\documentclass[twocolumn,floatfix,eqsecnum,prb,superscriptaddress,longbibliography]{revtex4-1}

\usepackage[english]{babel}
\usepackage{graphicx, float}
\usepackage{dcolumn}
\usepackage{bm}
\usepackage{lipsum}
\usepackage{mathtools}
\usepackage{braket,slashed}
\usepackage[caption=false]{subfig}
\usepackage{xcolor}
\usepackage{amssymb}
\usepackage{ulem}

\usepackage{hyperref}

\renewcommand{\selectlanguage}[1]{}
\begin{document}


\title{Berezinskii--Kosterlitz--Thouless transitions in a \\topological superconductor--ferromagnet--superconductor junction}

\author{Adrian Reich} \email{adrian.reich@kit.edu}
\affiliation{Institute for Theoretical Condensed Matter Physics, Karlsruhe Institute of Technology, 76131 Karlsruhe, Germany}

\author{Dmitriy S. Shapiro}
\affiliation{Institute for Quantum Materials and Technologies, Karlsruhe Institute of Technology, 76344 Eggenstein-Leopoldshafen, Germany}
\affiliation{Institute for Quantum Computing Analytics (PGI-12), Forschungszentrum J\"ulich, 52425 J\"ulich, Germany}

\author{Alexander Shnirman}
\affiliation{Institute for Theoretical Condensed Matter Physics, Karlsruhe Institute of Technology, 76131 Karlsruhe, Germany} 
\affiliation{Institute for Quantum Materials and Technologies, Karlsruhe Institute of Technology, 76344 Eggenstein-Leopoldshafen, Germany}

\date{\today}

\begin{abstract}

We investigate quantum phase transitions in a topological Josephson junction with an embedded ferromagnetic layer, revealing a rich landscape of critical phenomena. The low-energy excitations comprise Majorana fermions propagating along the junction, coupled to the magnons in the ferromagnet. Based on mean-field and renormalization group arguments, we predict Berezinskii-Kosterlitz-Thouless (BKT) transitions in this system, both in the case of a magnetic easy-plane and weak easy-axis anisotropy. In the latter case, this is based on an emergent effective easy-plane, spanned by the easy-axis and the component of the magnetization which couples to the Majoranas. We conclude by presenting a conjecture for the full phase diagram of the model. It covers BKT transitions as well as exotic multicritical and supersymmetric points known from related models of interacting real fermions and bosons.

\end{abstract}

\maketitle


\section{Introduction}

{\color{black}During the ongoing efforts to study and control Majorana modes in condensed matter systems~\cite{RevModPhys.83.1057, alicea_new_2012, Beenakker-Rev}, a variety of them -- e.g. topological insulator--superconductor structures~\cite{menard2017two,doi:10.1126/sciadv.aav6600}, quantum anomalous Hall insulator films~\cite{HeScience2017,shen2018spectroscopic}, Kitaev spin liquids~\cite{kasahara2018majorana},  and iron-based superconductors~\cite{Wang104} -- have proven to be promising platforms to host these exotic quasiparticles. Signatures of  Majorana modes were also claimed in the van der Waals CrBr$_3$/NbSe$_2$ heterostructure~\cite{Kezilebieke:2020ab}; however, these were later called into question because of a topologically trivial Yu-Shiba-Rusinov state reported in the same system~\cite{Li2024}.}

In this work, we focus our interest on long Josephson junction geometries comprised of $s$-wave superconductors proximity-coupled to the surface states of a 3D topological insulator (TI). These structures have been shown to resemble junctions of spinless two-dimensional $p_x+ip_y$ superconductors, which are able to host Majorana modes\cite{fu_superconducting_2008, FuKanePrl2009}.

Depositing a ferromagnetic strip in such a junction opens possibilities of manipulating the associated Josephson current and quasiparticle states\cite{tanaka_manipulation_2009,zyuzin_josephson_2016,bobkova_magnetoelectrics_2016,amundsen_vortex_2018}. However, when studying the effect of a ferromagnet's magnetization on the electronic properties, one has to keep in mind that the magnetization itself is not perfectly rigid, but may exhibit dynamics, as described by the Landau-Lifshitz-Gilbert equation.

In particular, in Ref.~\onlinecite{reich_magnetization_2023} such a setup has been studied under the assumption of a strong magnetic easy-axis anisotropy. There, the authors found that the spatial and temporal stiffness of the magnetization can stabilize a phase with spontaneous $\mathbb{Z}_2$-symmetry breaking, in which the magnetization is tilted away from the easy-axis and the one-dimensional Majorana fermions consequently gapped out. Solitonic excitations in this phase carry zero-dimensional Majorana bound states.
A closely related system of an anti-ferromagnet coupled to the edge of a $p_x+ip_y$ superconductor, known since as Grover-Sheng-Vishwanath model, was analyzed numerically in Ref.~\onlinecite{grover_emergent_2014}. The phase transition in this model was identified to belong to the supersymmetric tricritical Ising universality class.

The present work proposes an approach that encompasses arbitrary magnetic anisotropies. It predicts novel quantum phases in the easy-plane and weak easy-axis regime while also reproducing the strong easy-axis limit addressed in Ref.~\onlinecite{reich_magnetization_2023}. To this end, we derive an effective field theory from the full model for each case, motivated by examining the corresponding mean-field picture. Performing a momentum-shell renormalization group (RG) analysis, we identify Berezinskii-Kosterlitz-Thouless (BKT) transitions of the system and find that the magnetic stiffnesses are responsible for stabilizing any (quasi) long-range order.

We conclude with a conjecture for the complete phase diagram of the model, including both the (strong-coupling) tricritical Ising and (weak-coupling) BKT behavior in the easy-axis regime, as well as possible novel multicritical points.

\section{The model}

\begin{figure}
   \centering
   \includegraphics[width=0.5\textwidth]{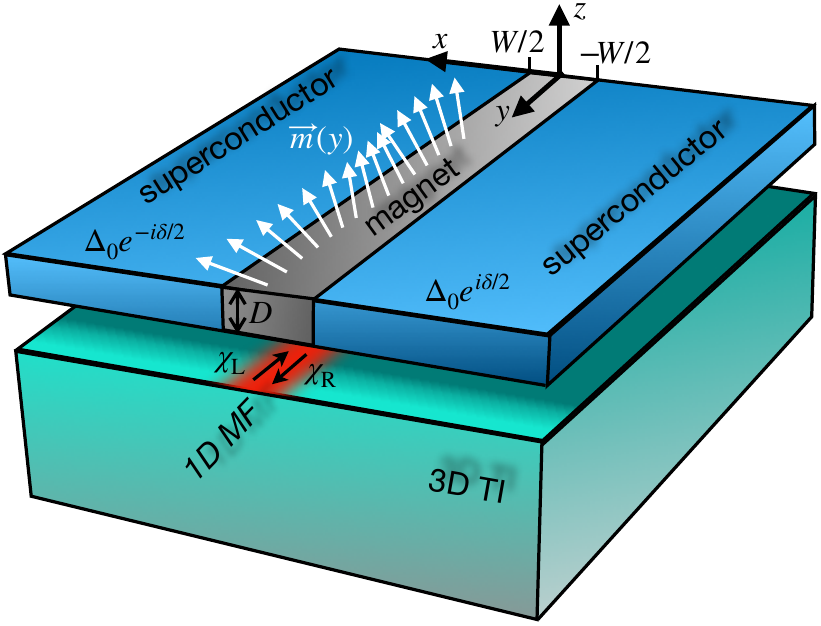}
 \caption{Considered geometry of a superconductor-ferromagnet-superconductor (SMS) junction of width $W$ {\color{black} and thickness $D$} on the surface of a 3D strong topological insulator (TI) with a phase difference $\delta$ between the superconductors. The low-energy physics of the 1D junction is described by counter-propagating Majorana fermions (MF) $\chi_{\rm R/L}$ which are coupled by the dynamic magnetization $\vec{M}(y)=|\vec{M}|\cdot\vec{m}(y)$.}
   \label{fig:junction}
\end{figure}

We consider the surface of a three-dimensional strong TI, covered by a superconductor-ferromagnet-superconductor (SMS) junction (see Fig.~\ref{fig:junction}). The TI surface states can be described by a single Dirac cone with Fermi velocity $v_{\rm F}$ and a chemical potential $\mu$. Due to the proximity effect, the $s$-wave superconductors induce superconducting gaps $\Delta_0e^{\pm i\delta/2}$, where the sign of the phase $\pm$ refers to the left-/right-hand side of the junction, respectively. The magnetization $\Vec{M}$ of the ferromagnetic insulator induces an effective exchange field $\vec{h}_{\rm eff} = \alpha\Vec{M}$ in the surface of the TI, coupling to the electronic spins.

In Ref.~\onlinecite{reich_magnetization_2023} it has been shown, akin to the seminal work by Fu and Kane\cite{fu_superconducting_2008}, that by treating the magnetization term as a small perturbation for a fixed phase difference between the superconductors $\delta=\pi$, the low-energy physics of the system can be described by two counter-propagating Majorana modes $\chi_{\rm R/L}$ in the junction, hybridized by the $x$-component of the magnetization. The corresponding Hamiltonian reads
\begin{equation}
    H_\chi = \int dy \Big[-\frac{iv}{2}(\chi_{\rm R}\partial_y\chi_{\rm R}-\chi_{\rm L}\partial_y\chi_{\rm L})+igm_x\chi_{\rm R}\chi_{\rm L}\Big],
\end{equation}
with $\Vec{m} = \vec{M}/|\vec{M}|$, coupling constant $g \simeq\alpha |\vec{M}|\frac{W}{v_\text{F}/\Delta_0}$ and effective fermionic velocity {\color{black} $v \simeq \frac{\Delta_0^2}{\mu^2}\big|\cos\left(\mu W/v_F\right)\big|\,v_{\rm F}$ in the experimentally relevant regime $\mu\gg \alpha|\vec{M}|,\Delta_0$.}

Additionally, we introduce a micromagnetic description of the magnetization dynamics with an exchange coupling $A$ and scalar anisotropy $B$ in $z$-direction (easy-axis if $B>0$, easy-plane if $B<0$). Under the assumption of sufficiently small width $W$, such that $\Vec{m}(\bm{r})\simeq\vec{m}(y)$, the magnetization dynamics can be described by the Lagrangian
\begin{equation}
\begin{split}
    &\mathcal{L}_m = M\dot{\varphi}(1-\cos\theta)\\
    &\hspace{1cm}-\frac{A}{2}\left[(\theta^\prime)^2+(\varphi^\prime)^2\cos^2\theta\right]+\frac{B}{2}\cos^2\theta,
\end{split}
\end{equation}
where the angles $(\theta(y),\varphi(y))$ parameterize the magnetization direction in spherical coordinates $\Vec{m}=(\sin\theta\cos\varphi, \sin\theta\sin\varphi,\cos\theta)^T$. In order for the equation of motion corresponding to $\mathcal{L}_m$ to be the dissipationless Landau-Lifshitz-Gilbert equation, we fix $M = |\Vec{M}|/\gamma$ with gyromagnetic ratio $\gamma$.

Altogether, the effective Euclidean action we are going to study reads
\begin{equation} \label{eq:action}
\begin{split}
    S = \int d\tau\,dy\left[\frac{v}{2}\bar\chi\slashed\partial\chi-iM(\partial_\tau\varphi)(1-\cos\theta)\right.&\\
    +\frac{A}{2}\left((\partial_y\theta)^2+(\partial_y\varphi)^2\cos^2\theta\right)-\frac{B}{2}\cos^2\theta&\\
    \hspace{4cm}\left.+\frac{g}{2}\bar\chi\chi\sin\theta\cos\varphi\right]&,
    \end{split}
\end{equation}

with the two-component Majorana spinor $\chi = (\chi_{\rm R},\chi_{\rm L})^T$, $\bar\chi=\chi^T\gamma_0$, the Dirac matrices $\gamma_0 = \sigma_y$ and $\gamma_1=\sigma_x$, as well as $\slashed\partial = \frac{1}{v}\gamma_0\partial_\tau+\gamma_1\partial_y$.

\section{Recap of results for strong easy-axis anisotropy \label{sec:strong_easy-axis}}

\begin{figure}
   \centering
   \includegraphics[width=.5\textwidth]{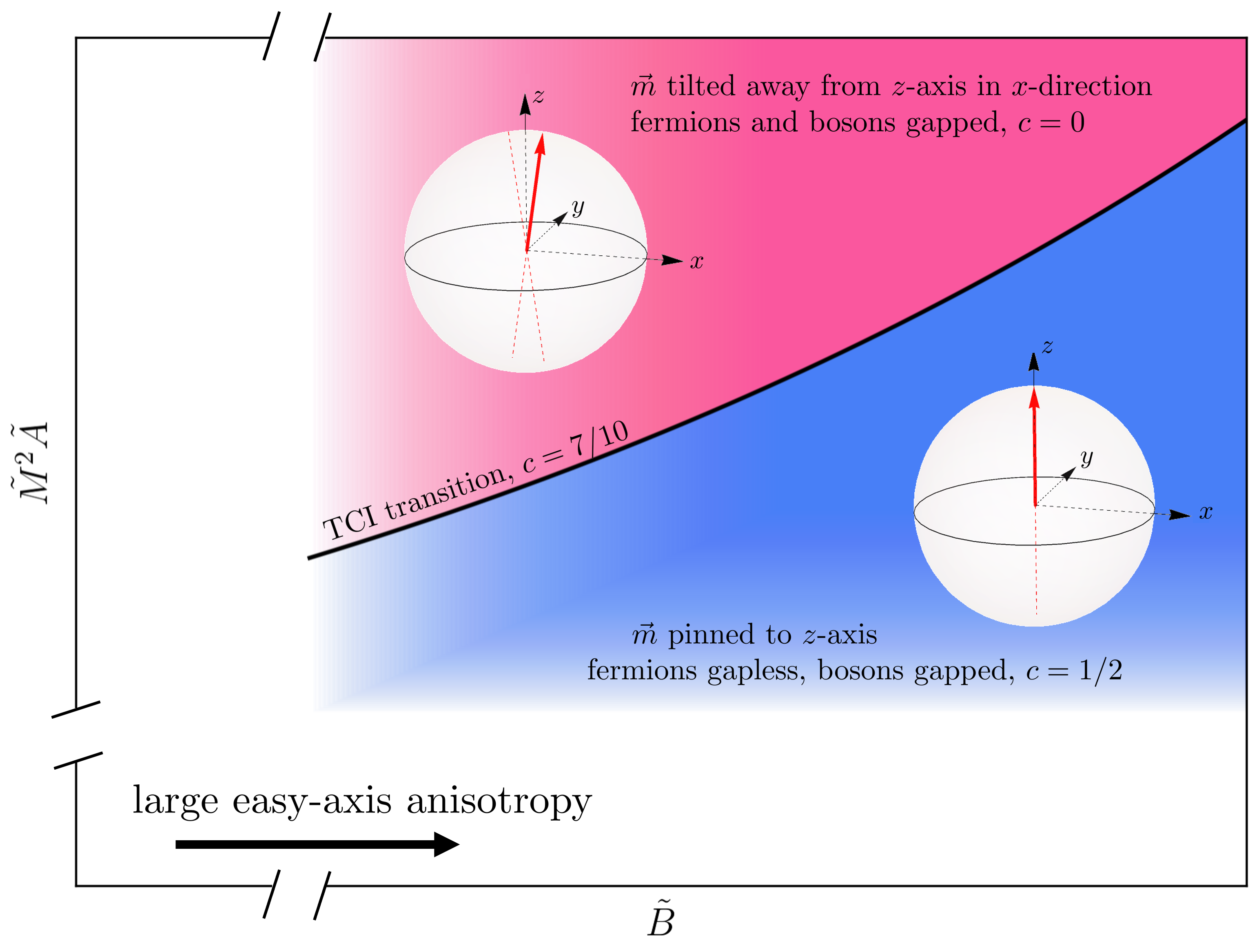}
   \caption{Schematic phase diagram for the case of a strong easy-axis anisotropy $B$. The Peierls-like instability leading to a tilting of the magnetization away from the (easy-)$z$-axis due to the magnon-Majorana interaction is stabilized by the stiffnesses $M$ and $A$.}
   \label{fig:phase-diagram-TCI}
\end{figure}

In Ref.~\onlinecite{reich_magnetization_2023}, the case of a strong easy-axis anisotropy $B>0$ for this model was considered, such that $m_x,m_y\ll 1$ throughout. In this regime, the authors found a quantum phase transition between a massive $\braket{m_x}\neq 0$ and a massless $\braket{m_x}=0$ phase with critical Majorana fermions (central charge $c=1/2$) to take place at sufficiently large $M$ and/or $A$. The corresponding spontaneous symmetry breaking, leading to a nonzero vacuum expectation value of $m_x$ in the massive phase, is in mean-field theory found to be of Peierls type $\braket{m_x}_{\rm MF}\sim e^{-vB/g^2}$. The parameters $M$ and $A$ act as stiffnesses of the magnetization dynamics in time and space, respectively, stabilizing the mean-field result and thus the broken symmetry against quantum fluctuations. The larger the anisotropy, and thus the smaller $\braket{m_x}_{\rm MF}$, the larger the critical stiffnesses were found to be. The resulting phase diagram is sketched in Fig.~\ref{fig:phase-diagram-TCI}.

Equivalently, one can think of $M$ and $A$ as controlling the nonlocality of an effective 4-Majorana interaction. The corresponding correlation lengths of this interaction read $\xi_x \sim \sqrt{A/B}$ in space and $\xi_\tau \sim M/B$ in time. The interaction can be shown to be RG irrelevant in the local case, i.e. if $\xi_x,v\xi_\tau\lesssim \Lambda^{-1}$ with some short-distance cut-off $\Lambda^{-1}$, such that only sufficiently large values of $A$ and $M$ enable the possibility of a phase transition at strong coupling. 

Numerical analyses of similar models\cite{grover_emergent_2014,rahmani_emergent_2015} suggest the corresponding phase transition to belong to the tricritical Ising (TCI) universality class with $c=7/10$. 
{\color{black} This can be understood as a consequence of the Majorana character of the fermions, since a theory of free Majoranas is equivalent to the critical Ising phase with central charge $c=1/2$. The TCI class is then the most natural way to transition from such a gapless phase to one with an Ising-like broken symmetry. Note that domain walls in the gapped phase furthermore carry the coveted Majorana zero modes.}

\section{Easy-plane anisotropy \label{sec:easy-plane}}

In contrast to the case discussed above, in the following we are going to consider an easy-plane anisotropy with $B<0$, before returning to the case $B>0$ in section \ref{sec:easy-axis}.

One-dimensional ferromagnetic chains with a sufficiently large easy-plane anisotropy are known to be well described by an $XY$-type model\cite{schollwock_quantum_2004}. If the continuous symmetry in the easy-plane is broken by means of a Zeeman field, the long-wavelength dynamics can be mapped to a sine-Gordon action for the azimuthal angle $\varphi$ and the Zeeman field can thus induce a BKT transition\cite{schollwock_quantum_2004,mikeska_solitons_1977}. In the following, we will see that the Majorana-magnon interaction in our system plays a very similar role to such a Zeeman field by breaking the continuous symmetry and pinning the magnetization. 

\subsection{Mean-field theory}

\begin{figure*}
    \centering
    \includegraphics[width=\linewidth]{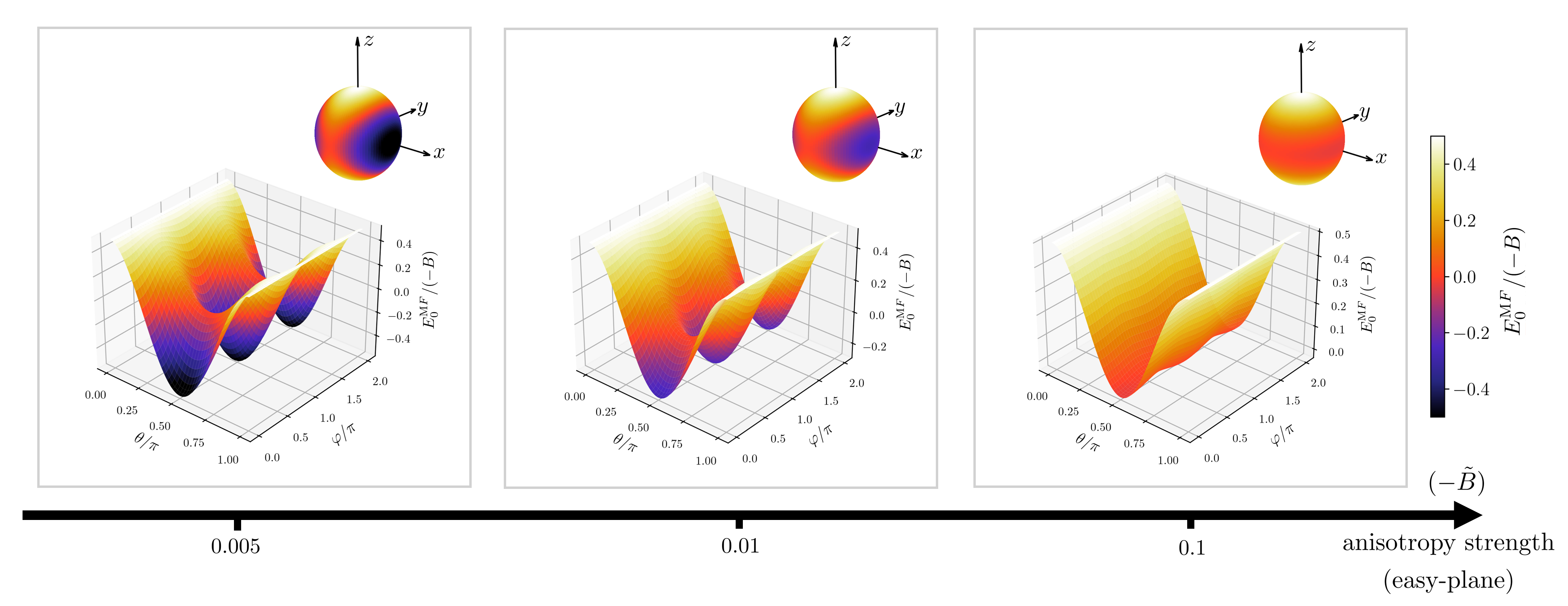}
    \caption{Plots of the mean-field ground state energy $E_0^{\rm MF}(\theta,\varphi)$, given by (\ref{eq:E_MF-easyplane}), divided by the easy-plane anisotropy strength $(-B)$ for a fixed value of the coupling constant $\tilde{g}^2 = 0.01$. The small diagrams show the projection onto the unit sphere.}
    \label{fig:mf_easy-plane}
\end{figure*}

Integrating out the fermions from (\ref{eq:action}) and for now simply assuming $\theta$ and $\varphi$ to be constant, one finds the mean-field ground state energy density $E^{\rm MF}_0$ to read
\begin{equation} \label{eq:E_MF-easyplane}
\begin{split}
    &\frac{1}{v\Lambda^2}E^{\rm MF}_0(\theta,\varphi) = -\frac{\Tilde{B}}{2}\cos^2\theta\ +\\ &+\frac{\Tilde{g}^2}{8\pi}\sin^2\theta\cos^2\varphi\left\{\log\left(\frac{\Tilde{g}^2}{4}\sin^2\theta\cos^2\varphi\right)-1\right\},
\end{split}
\end{equation}
where we introduced the dimensionless parameters $\Tilde{B}=B/(v\Lambda^2)$ and $\Tilde{g}=g/(v\Lambda)$. The resulting energy landscape for different values of $(-\Tilde{B})$ and $\Tilde{g}$ is plotted in Fig.~\ref{fig:mf_easy-plane}.

For large values of $(-\Tilde{B}) \gg \Tilde{g}^2$, the first term in (\ref{eq:E_MF-easyplane}) dominates and the situation resembles the pure easy-plane picture one would expect: the energy is minimal for configurations with $\vec{m}$ in the $x$-$y$-plane, i.e. with $\theta = \pi/2$, and there is an approximate continuous symmetry regarding rotations around the $z$-axis, i.e. the energy does not depend on $\varphi$. 
This suggests the strong easy-plane case to correspond to a phase with $\varphi$ fluctuating freely, while $\theta$ is fixed to $\pi/2$, i.e. a pure $XY$ model. If in contrast $(-\Tilde{B})\ll\Tilde{g}^2$, the interaction with the Majorana fermions dominates and leads to pronounced minima on the $x$-axis at $\varphi=0$ and $\varphi=\pi$ (still with $\theta = \pi/2$). 
With regards to the action (\ref{eq:action}), this signals a fermionic gap opening with a spontaneous $\mathbb{Z}_2$ symmetry breaking akin to the strong easy-axis case discussed in Section \ref{sec:strong_easy-axis}. Note however, that there the mean field minima were exponentially close to each other, whereas here they lie on opposite ends of the unit sphere. In analogy to our earlier discussion, we then expect the stability against quantum fluctuations of this massive phase, emerging in the mean-field picture, to be dependent on sufficiently large values of the stiffnesses $A$ and/or $M$. The less pronounced the minima, i.e. the larger $(-\Tilde{B})$ compared to $\Tilde{g}^2$, the larger the $A$- and $M$-values required for the stability become. If the stiffnesses are too small, the minima get smeared out and the system resides in a massless phase with free $\varphi$.

Note that we operate here and below under the assumption, that even the ``small'' stiffnesses are large enough to sufficiently suppress fluctuations in $\theta$-direction, i.e. fluctuations out of the easy-plane, which is reasonable for a ferromagnet. 

\subsection{Effective theory near $x$-$y$-plane and RG analysis}
\begin{figure*}[t]
    \centering
    \vspace{-.3cm}
    \subfloat[]{
        \includegraphics[width=.45\textwidth]{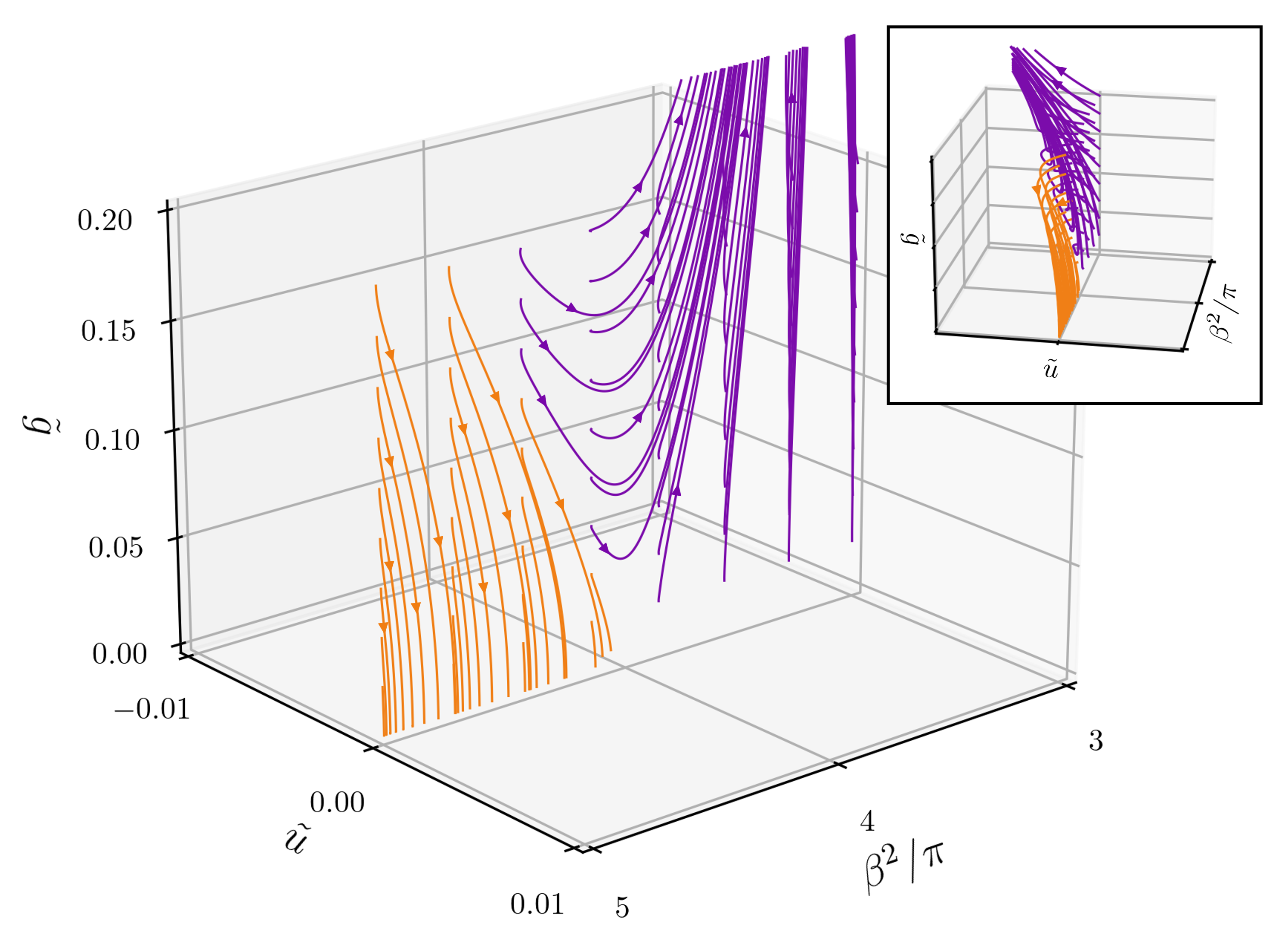}
        \label{fig:flow_easy-plane}}
        \hspace{1cm}
        \subfloat[]{
        \includegraphics[width=.45\textwidth]{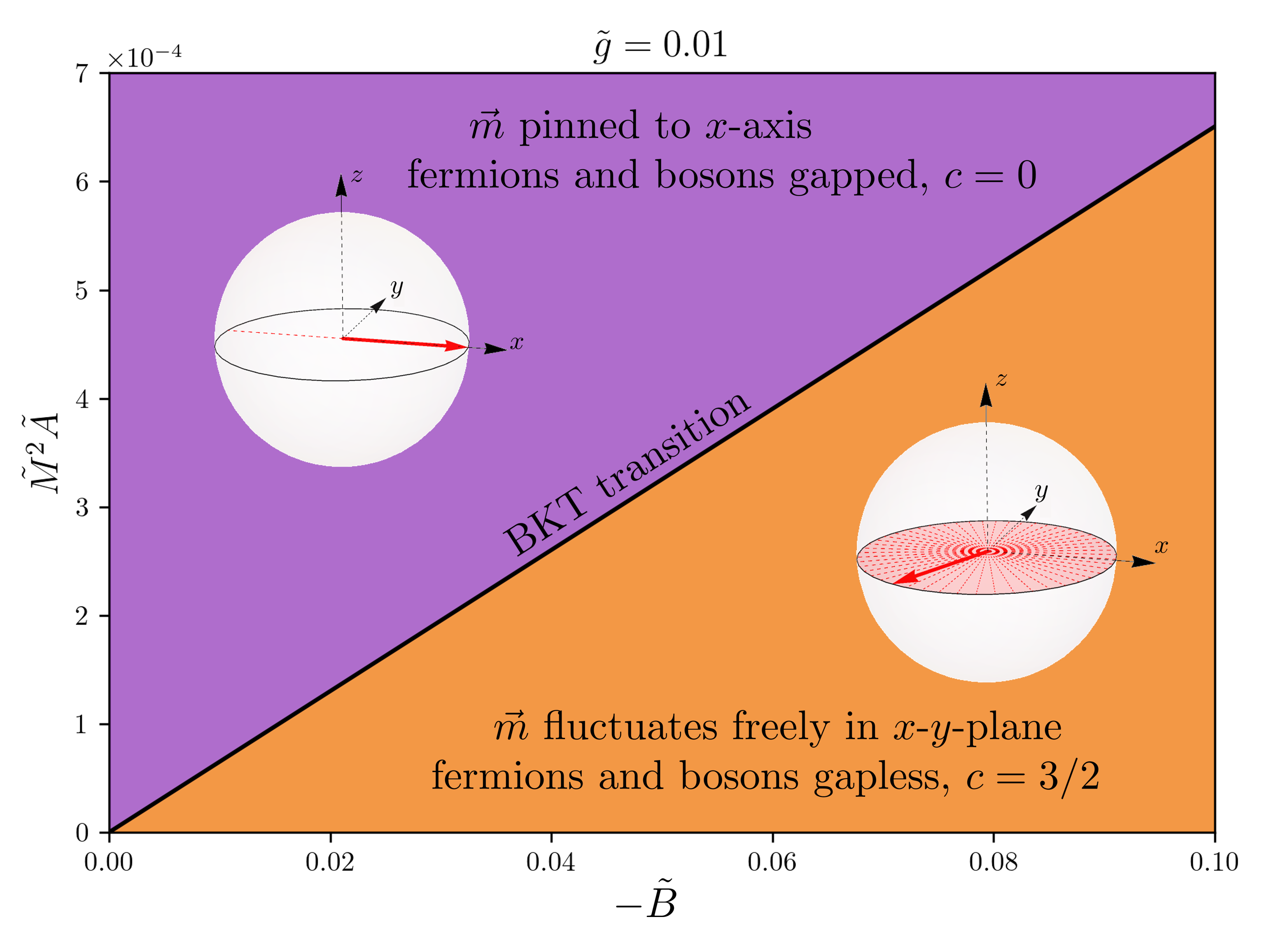}
        \label{fig:phasediagram_easy-plane_g=0.01}}
        
    \caption{(a) RG flow corresponding to Eqns.~(\ref{eq:flow-beta})-(\ref{eq:flow-g}) starting in the $\tilde{u}=0$-plane. The characteristic BKT flow with a transition from a regime with $\tilde{u},\tilde{g}\rightarrow 0$ (orange) to a strong-coupling regime (purple) is evident. The inset shows the same diagram from a rotated perspective to make the flow towards negative $\tilde{u}$ visible. (b) Resulting phase diagram in terms of the original parameters at $\tilde{g}=0.01$, showing increasing critical values of the stiffnesses for increasing anisotropy ($\tilde{A} = A/v, \tilde{M} = M/\Lambda$). The slope of the critical line diminishes when increasing the coupling constant $\tilde{g}$.}
    \label{fig:flow-and-phasediagrams_easy-plane}
\end{figure*}

In the strong easy-plane anisotropy case, $(-\Tilde{B})\gg\Tilde{g}^2$, we saw that it is  natural to limit the theory to the $x$-$y$-plane and only consider small fluctuations around it, $\theta \simeq \pi/2+\delta\theta$ with $\delta\theta\ll 1$, as is reflected in the mean-field results in Fig.~\ref{fig:mf_easy-plane}. Furthermore, the mean-field picture suggests that for the possible phase transition to a phase with massive fermions and the rotational $\varphi$-symmetry broken down to a spontaneously broken $\mathbb{Z}_2$-symmetry, only configurations in the $x$-$y$-plane are of importance as well. We therefore proceed by deriving an effective field theory valid in the vicinity of the $x$-$y$-plane.

Replacing $\theta = \pi/2 + \delta\theta$ in the action (\ref{eq:action}) and only keeping terms involving $\delta\theta$ up to Gaussian order, it follows
\begin{equation}
\begin{split}
    S &\simeq -iM\int d\tau\,dy\,(\partial_\tau\varphi) + \int d\tau\,dy\left[\frac{v}{2}\bar\chi\slashed\partial\chi+iM(\partial_\tau\varphi)\delta\theta
    \right.\\&+\frac{A}{2}\left((\partial_y\delta\theta)^2+(\partial_y\varphi)^2\right)
    \left.-\frac{B}{2}\delta\theta^2+\frac{g}{2}\bar\chi\chi\cos\varphi\right].
\end{split}
\end{equation}
The first term is a theta term, which is trivial in the present case and can be omitted. Integrating out $\delta\theta$ yields
\begin{align}
\begin{split}
    &S = \int d\tau\,dy\,\left[\frac{v}{2}\bar\chi\slashed\partial\chi + \frac{g}{2}\bar\chi\chi\cos\varphi\right] \\ 
    + &\frac{1}{2}\int\frac{d\omega\,dq}{(2\pi)^2}\varphi_{-q,-\omega}\left(\frac{M^2\omega^2}{A^\prime q^2-B}+Aq^2\right)\varphi_{q,\omega} \label{eq:action-with-Aprime}
\end{split}
\end{align}
where $\varphi(y,\tau) = \int\frac{d\omega\,dq}{(2\pi)^2}\,\varphi_{q,\omega}e^{iqy-i\omega\tau}$ and we introduced the new parameter $A^\prime$, which microscopically equals $A$ but scales differently under RG than the $A$ already present. In particular, we find at tree-level $dA^\prime/dl = -2A^\prime$. $A^\prime$ is thus strongly irrelevant and will be omitted in the following. In Appendix \ref{app:checkAprime}, we explicitly check that even if $A^\prime\Lambda^2 \gg (-B)$ holds microscopically, the $A^\prime$-term can safely be neglected from the outset without qualitatively altering the resulting phase diagrams. 

Additionally, we will see that a $\cos 2\varphi$-term is generated under RG flow and the resulting effective theory can thus be written as
\begin{align}
\begin{split}
    S = \int d\tau\,dy\,&\left[\frac{v}{2}\bar\chi\slashed\partial\chi + \frac{1}{2}\left(\frac{1}{c}(\partial_\tau\phi)^2+c(\partial_x\phi)^2\right)\right. \\
    &\ \ \left.+ u\cos 2\beta\phi + \frac{g}{2}\bar\chi\chi\cos\beta\phi\right]. \label{eq:effective-action}
\end{split}
\end{align}
Here, we defined $\beta^2 = \sqrt{-B/(M^2A)}$, the effective bosonic velocity $c = \sqrt{-BA/M^2}$ and the rescaled field $\phi = \varphi/\beta$. Microscopically, $u=0$.

This model is evocative of the supersymmetric sine-Gordon (SSG) model, which it corresponds to for $g^2 = -4\beta^2u$ and $u<0$. It is known that the SSG model always flows to a gapped phase\cite{witten_properties_1978}, which can be checked to be consistent with Eqns.~(\ref{eq:flow-beta})-(\ref{eq:flow-g}).

Employing a momentum-shell RG analysis, for which the details can be found in Appendix \ref{app:rg}, yields the flow equations (up to second order in $\tilde{u}=u/(v\Lambda^2)$ and $\tilde{g}$) below. They are valid as long as $\beta^2 > \pi/2$ and read
\begin{align}
    \frac{dc}{dl} &= \frac{\beta^2\tilde{g}^2}{16\pi v/c}\left(\frac{v^2}{c^2}-1\right)c, \label{eq:flow-c} \\
    \frac{d\beta^2}{dl} &= -2C_1\frac{v^2}{c^2}\beta^6\tilde{u}^2 - \frac{\beta^4\tilde{g}^2}{8\pi v/c}, \label{eq:flow-beta} \\ 
    \frac{d\tilde{u}}{dl} &= 2\tilde{u}\left(1-\frac{\beta^2}{2\pi}\right) - \frac{\tilde{g}^2}{8\pi}, \label{eq:flow-u} \\
    \frac{d\tilde{g}}{dl} &= \tilde{g}\left(1-\frac{\beta^2}{4\pi}-C_2\frac{v}{c}\beta^2\tilde{u}\right), \label{eq:flow-g}
\end{align}

where $C_{1/2}$ are numerical constants (see Appendix \ref{app:rg} for details). Similar flow equations for this type of model have been obtained in Ref.~\onlinecite{chen_notitle_2019}. 

From Eq.~(\ref{eq:flow-c}) follows an emergent Lorentz symmetry. The flow of the remaining three parameters, starting in the $\tilde{u}=0$-plane relevant for this problem, is shown in Fig.~\ref{fig:flow_easy-plane}. It exhibits the characteristics of a BKT transition: there is a line of fixed points on the $\beta^2$-axis, and a transition near $\beta^2=4\pi$, from a regime with the couplings running to zero (orange) to a strong coupling regime with $\tilde{g}\rightarrow\infty$ and $\tilde{u}\rightarrow -\infty$ (purple).

This corresponds to a phase transition from unbounded fluctuations of $\varphi$, being described by a free massless theory, to a massive phase with $\braket{\varphi} = 0$ or $\pi$, i.e. the magnetization being pinned to the $x$-axis, spontaneously breaking the inversion symmetry. An expectation value $\braket{\varphi} \neq \pm \pi /2$, and thus $\braket{m_x}\neq 0$, gaps out the Majorana fermions.

Plotting the resulting phase diagram in terms of the original parameters of the problem in Fig.~\ref{fig:phasediagram_easy-plane_g=0.01}, reveals that the RG considerations confirm the mean-field picture: the larger the easy-plane anisotropy $(-B)$, the larger the stiffnesses $A$ and/or $M$ need to be in order to stabilize the massive phase.

\section{Easy-axis anisotropy \label{sec:easy-axis}}

In section \ref{sec:strong_easy-axis}, we recapped results of an earlier work for the case of an easy-axis anisotropy $B>0$ of such a magnitude, that only configurations of the magnetization with $m_x,m_y\ll 1$, i.e. near the $z$-axis, were of interest. For weaker easy-axis anisotropies, no such immediate simplifications of the problem are obvious. Still, in the following, we are going to attempt a generalization to weaker easy-axis anisotropies based on observations within mean-field theory: there, as we will see below, a regime can be identified in which an effective easy-plane is spanned by the easy-axis and the axis perpendicular to the junction, along which the interaction with the Majoranas takes place. This allows us, using similar arguments as above, to again postulate an effective field theory for which an RG analysis can be performed.

We would like to stress, however, that, compared to the previous section, the arguments that lead to the effective theory here are of rather conjectural character. Still, we interpret the consistency between the mean-field and RG results to be an indication of the merit of this approach as a step towards an understanding of the full phase diagram.

\subsection{Mean-field theory}

\begin{figure*}
    \centering
    \includegraphics[width=\linewidth]{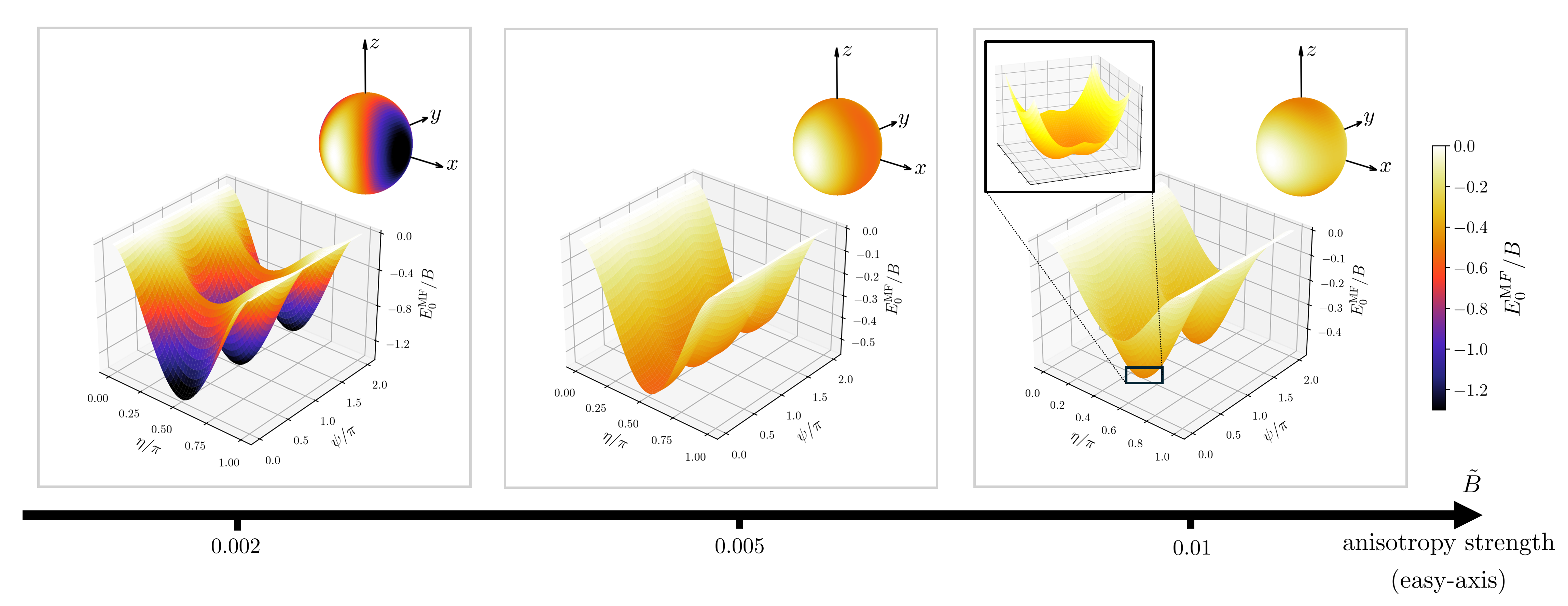}
    \caption{Plots of the mean-field ground state energy $E_0^{\rm MF}(\eta,\psi)$ divided by the easy-axis anisotropy strength $B$ for a fixed value of the coupling constant $\tilde{g}^2 = 0.01$. The small diagrams show the projection onto the unit sphere.}
    \label{fig:mf_easy-axis}
\end{figure*}

\begin{figure*}[t]
    \centering
    \subfloat[]{
        \includegraphics[width=.45\textwidth]{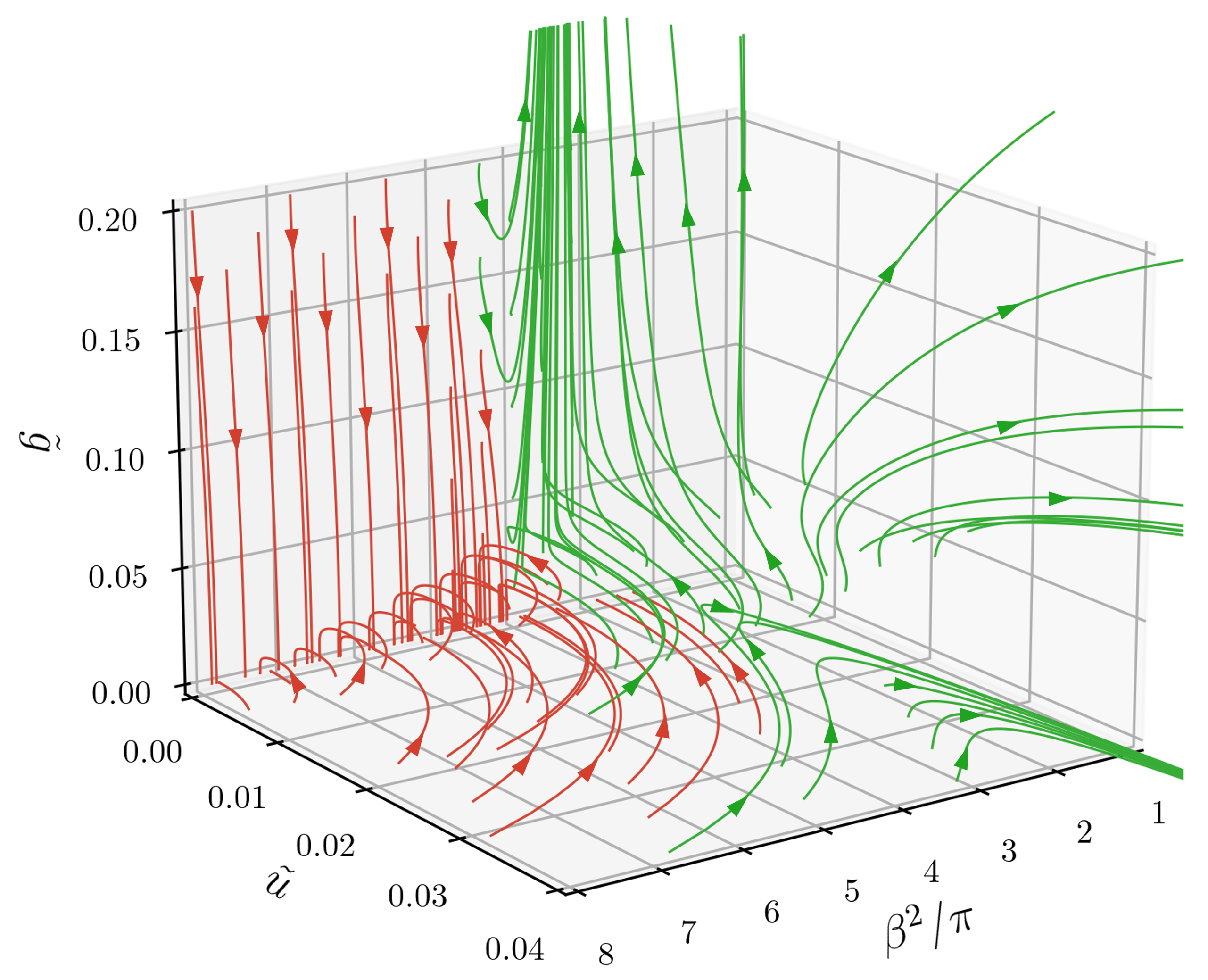}
        \label{fig:flow_easy-axis}}
    \subfloat[]{
        \includegraphics[width=.45\textwidth]{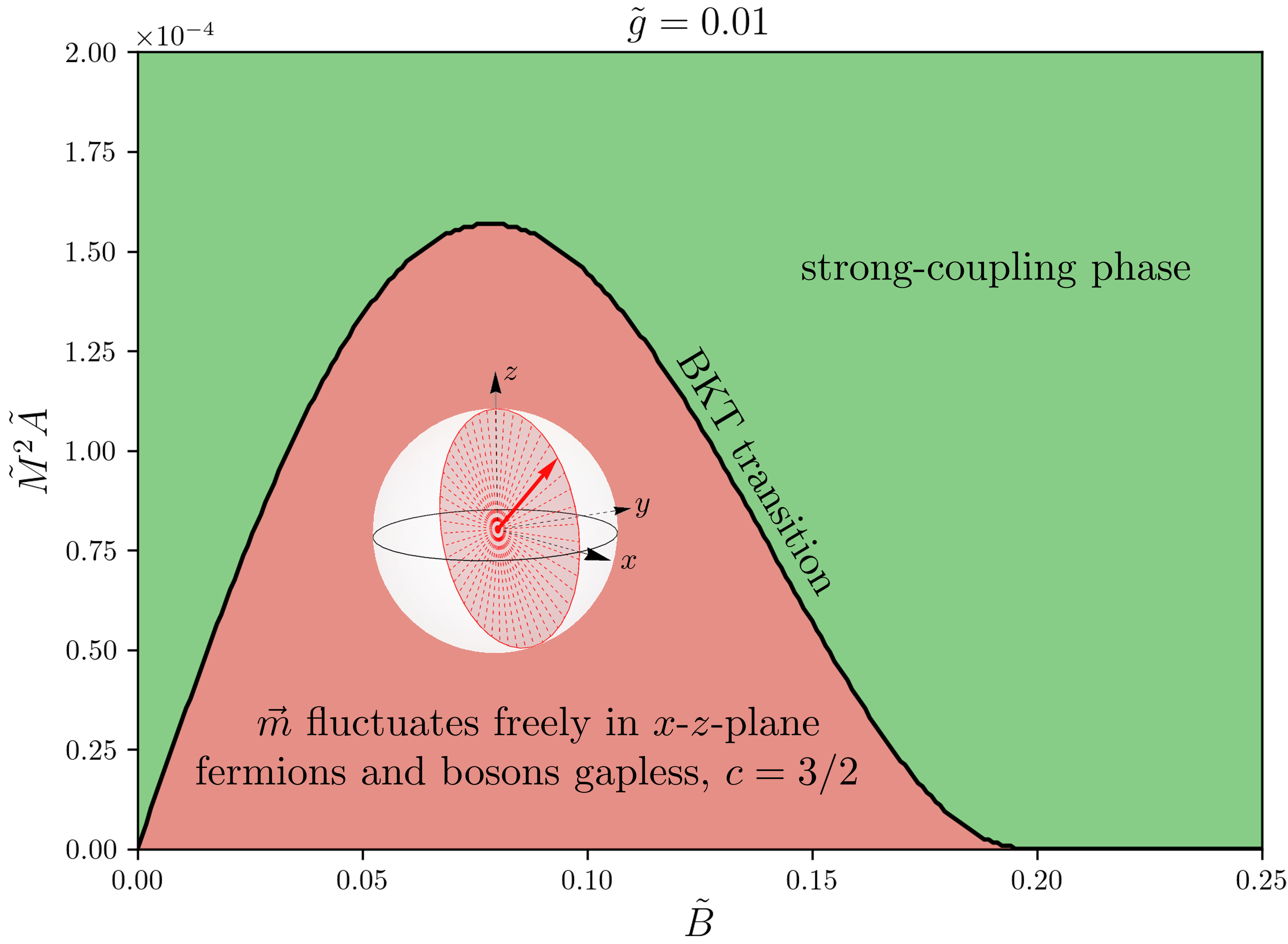}
        \label{fig:phasediagram_easy-axis_g=0.01}
        }
    \caption{(a) RG flow corresponding to Eqns.~(\ref{eq:flow-beta})-(\ref{eq:flow-g}). As compared to Fig.~\ref{fig:flow_easy-plane}, an additional BKT transition to a different strong coupling regime ($\tilde{u}\rightarrow\infty, \tilde{g}\rightarrow 0$) emerges, most clearly visible in the vicinity of the $\tilde{g}=0$-plane. The interpolation between the BKT transitions in the $\tilde{u}=0$- and the $\tilde{g}=0$-plane forms a critical surface, separating the different strong-coupling regimes (green) from the one with $\tilde{u},\tilde{g}\rightarrow 0$ (red). (b) Resulting phase diagram in terms of the original parameters at $\tilde{g}=0.01$, showing a non-monotonous behaviour of the critical values of the stiffnesses as a function of the anisotropy. This can be linked to the observation that in mean-field theory the energy landscape becomes shallowest and thus allows for the largest fluctuations at some finite value of $B$.}
    \label{fig:flow-and-phasediagrams_easy-axis}
\end{figure*}

In the easy-axis case $B>0$, configurations with $\vec{m}$ pointing in $z$-direction will be important. In the parametrization we used until now, this is problematic, as $\varphi$ is not well defined if $\theta = 0$ or $\pi$. Let us therefore introduce new angles $(\eta,\psi)$ with which $\Vec{m}=(\sin\eta\cos\psi,-\cos\eta,\sin\eta\sin\psi)^T$. The resulting mean-field ground state energy density is plotted in Fig.~\ref{fig:mf_easy-axis} for different values of $\tilde{B}$. 
Analogously to the easy-plane case, at weak anisotropies $\tilde{B}\ll\tilde{g}^2$ the interaction dominates and leads to minima near the $x$-axis, whereas at $\tilde{B}\gtrsim\tilde{g}^2$ the regime discussed in section \ref{sec:strong_easy-axis} with minima near the $z$-axis is recovered. For $\tilde{B}\sim\tilde{g}^2/2$, the system resides in a transitional regime between these two edge cases, as the minima, stemming respectively from interaction and anisotropy, merge. 
Remarkably, this results in an energy landscape which resembles an easy-plane anisotropy, as the energy is minimal and nearly degenerate for all $\psi$ at $\eta=\pi/2$. Effectively, in this regime one can think of an easy-plane being spanned by the easy-axis-direction and the magnon-Majorana interaction. Mean-field theory thus suggests that here only configurations in the $x$-$z$-plane are of importance, as furthermore the energy barriers for fluctuations in $\eta$-direction are always larger than for the ones in $\psi$-direction. In analogy to our approach of the easy-plane case, we therefore suggest the low-energy physics to be reasonably well captured by an effective theory for $\psi$ which is valid near $\eta=\pi/2$.

\subsection{Effective theory near $x$-$z$-plane and RG analysis}
\begin{figure*}[t]
    \centering
    \includegraphics[width=\textwidth]{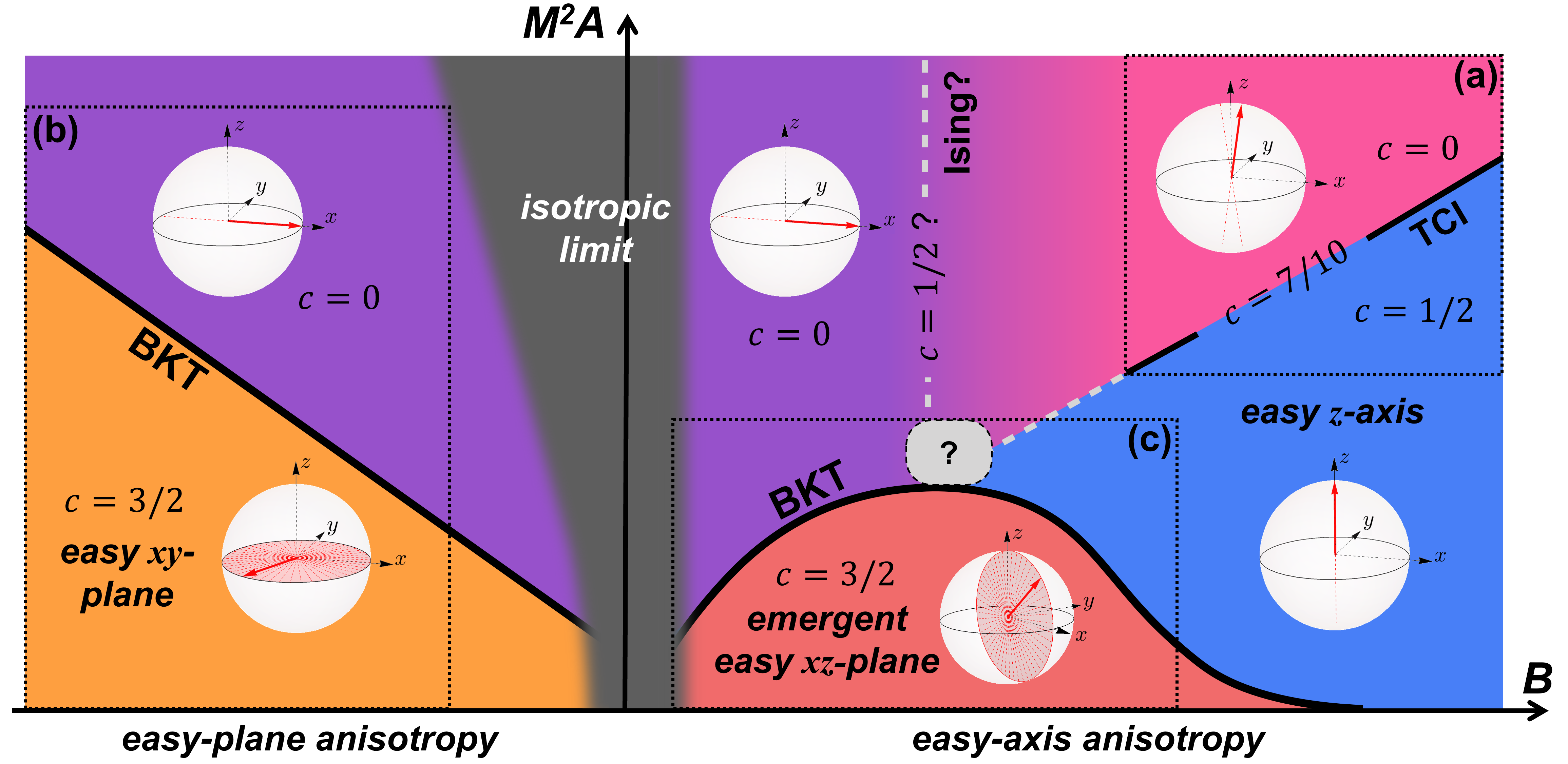}
    \caption{Conjecture for the full phase diagram at a fixed, finite $g$. The dotted boxes mark the areas for which we conducted an analysis in this paper, the areas outside of that are a subject of speculation. The TCI transition in the dotted box (a) is discussed in Section \ref{sec:strong_easy-axis}, the BKT transitions in the dotted boxes (b) and (c) are analyzed and discussed in Sections \ref{sec:easy-plane} and \ref{sec:easy-axis}, respectively. The (green) strong coupling phase of Fig.~\ref{fig:phasediagram_easy-axis_g=0.01} for $B>0$ is here comprised of the phases colored blue, pink and purple. The speculations of Section \ref{sec:conjecture} furthermore lead us to expecting an Ising transition to take place between a phase with both $x$- and $z$-inversion symmetry spontaneously broken (pink) and one with $x$-inversion spontaneously broken (purple). The way in which the Ising and the TCI line connect to each other and/or the BKT line remains an open question. The isotropic limit $B\rightarrow 0$ (dark grey) is out of the scope of this work but likely characterized by strong fluctuations.}
    \label{fig:full-phase-diagram}
\end{figure*}

Motivated by mean-field theory, we thus only consider small fluctuations out of the $x$-$z$-plane and take $\eta\simeq \pi/2+\delta\eta$ with $\delta\eta\ll 1$. Once again, only keeping terms involving $\delta\eta$ up to Gaussian order, subsequently integrating out $\delta\eta$ and omitting the irrelevant $A^\prime$-contribution, results in an action of the form (\ref{eq:effective-action}),
\begin{align}
\begin{split}
    S = \int d\tau\,dy\,&\left[\frac{v}{2}\bar\chi\slashed\partial\chi + \frac{1}{2}\left(\frac{1}{c}(\partial_\tau\Psi)^2+c(\partial_x\Psi)^2\right)\right. \\
    &\ \ \left.+ u\cos 2\beta\Psi + \frac{g}{2}\bar\chi\chi\cos\beta\Psi\right], \label{eq:effective-action-psi}
\end{split}
\end{align}
where now $\beta^2=\sqrt{B/(2M^2A)}$, $c=\sqrt{AB/(2M^2)}$, $u=B/4$ and $\Psi = \psi/\beta$. Crucially, even though this is of course the same theory with the same flow equations \eqref{eq:flow-c}-\eqref{eq:flow-g} as in the easy-plane case, here $u \neq 0$ microscopically, which results in a more involved phase structure. Specifically, note that positive values of $\tilde{u}$ slow and may even revert the growth of $\tilde{g}$. 
If the initial value of $\tilde{u}$ is sufficiently large, the flow might lead into a different strong-coupling regime than the one above (which is still present here, of course), namely one with $\tilde{u}\rightarrow\infty$ and $\tilde{g}=0$, see Fig.~\ref{fig:flow_easy-axis}. This would hint at $\braket{\psi}=\pm\pi/2$, i.e. the magnetization pointing in $z$-direction, corresponding to the regime with critical Majorana fermions discussed in section \ref{sec:strong_easy-axis}. 
In between these two possible strong-coupling limits, $\tilde{u}$ and $\tilde{g}$ are growing in competition with each other. Rigorously differentiating between the strong-coupling regimes (green in Fig.~\ref{fig:flow_easy-axis}) is not possible in this framework, on account of the derived flow equations only being valid for $\beta^2>\pi/2$ and the result of a weak-coupling expansion. Still, the BKT transitions near the $\beta^2$-axis are well controlled. This allows the identification of a massless phase (red), where $\tilde{u},\tilde{g}\rightarrow 0$ with $\psi$ fluctuating without bound. 

Looking at the resulting phase diagram at fixed $g$ in the $M^2A$-$B$-plane shows a non-monotonous behaviour of the critical value of the stiffnesses with a maximum at finite $B$ (see Fig.~\ref{fig:phasediagram_easy-axis_g=0.01}). This is in agreement with what can be expected from the mean-field picture: the energy landscape is the shallowest in the transitional regime between weak and strong anisotropy, which is why there the largest values for the stiffnesses are necessary in order to stabilize a phase with $\braket{\psi}\neq 0$.

{\color{black}
\subsection{Experimental accessibility}

Our analysis is primarily focused on qualitative aspects of the quantum phase structure of the given model. Still, for the sake of completeness, we proceed by roughly estimating the experimental accessibility of the identified relevant parameter regimes.

In Refs.~\onlinecite{mogi2019large,watanabe2019}, evidence for magnetic exchange gaps $\alpha |\vec{M}|$ induced in TI surface states via covering with a ferromagnetic insulator as large as several tens meV has been provided. Assuming a superconducting gap corresponding to a few Kelvins as the high-energy cut-off $v\Lambda = \Delta_0\sim 0.1-1\,$meV, we thus see that the coupling constant $g\sim \alpha |\vec{M}| \frac{W}{\xi}$ can be on the order of the cut-off even for very conservative estimates of the magnetic exchange coupling and/or very narrow junctions. The small values of $g/\Delta_0$, needed for our analysis to hold true, can therefore be presumed to be readily attainable.

In order to estimate the regions which are accessible in the $\tilde{M}^2\tilde{A}$-$\tilde{B}$-phase diagram with realistic materials, consider for example the 3D TI Bi$_2$Se$_3$. Its surface states exhibit a Fermi energy of $\mu \simeq 300\,$meV and a Fermi velocity of $v_{\text F}\simeq 400\,$meV\,nm \cite{xia_observation_2009,zhang_topological_2009}. The corresponding superconducting coherence length is $\xi = v_{\text F}/\Delta_0 \sim 400-4000\,$nm. Values for the micromagnetic parameters of ferromagnetic insulators (see e.g.\ Refs.~\onlinecite{POPOVA2013139,Zhang_2016,talalaevskij2017magnetic,mogi2018,tian2021dyocl}) typically fall in the range of $A/(DW) \sim 10^{-12}-10^{-11}\,$Jm$^{-1}$ for the exchange stiffness, $B/(DW) \sim 10^3-10^6\,$Jm$^{-3}$ for the scalar anisotropy constant as well as $|\vec{M}|/(DW) \sim 10^{4}-10^{5}\,$Am$^{-1}$ for the saturation magnetization, which leads to a ratio between saturation magnetization and gyromagnetic ratio ($\gamma\sim 10^{11}\,$s$^{-1}$T$^{-1}$) of $M/(DW) \sim 10^{27}-10^{28}\hbar\,$m$^{-3}$. Here, $D$ is the thickness of the magnetic covering. The characteristic length scale over which the magnetization varies is given by the magnetic exchange length $l_\text{ex} = \sqrt{2A/\mu_0|\vec{M}|^2}\sim 10-400\,$nm. As indicated in the listed references, these properties are generally not purely material-specific, but depend on the geometry and temperature of the sample.

The values of the dimensionless parameters we are interested in are given by $\tilde{B} = vB/\Delta_0^2$ as well as $\tilde{M}^2\tilde{A} = vM^2A/\Delta_0^2$. They are thus proportional to $v\propto |\cos(\mu W/v_{\rm F})|$.

With the given estimates, we find experimentally viable parameters, depending on the width $W$ and thickness $D$ of the magnetic film, to read $\tilde{B} \simeq 10^{-5}-10^{-2} \,\frac{DW}{\rm{nm}^2}|\cos\left(\mu W/v_F\right)|$, $\tilde{M}^2\tilde{A} \simeq 10^{-2}-10^1\,\left(\frac{DW}{\rm{nm}^2}\right)^3|\cos\left(\mu W/v_F\right)|$.

Consequently, due to the large stiffnesses naturally associated with ferromagnets, in order to roughly land in the region of the phase diagram that is examined in Fig.~\ref{fig:flow-and-phasediagrams_easy-axis}, a small cross sectional area with e.g. $D\sim W \sim 0.6-0.8\,$nm is necessary, which corresponds to few atomic layers in each direction. While typically the dimensions of magnetic thin films for applications like spintronics range from a few nanometers to a few micrometers, atomically thin layers exhibiting ferromagnetism have been succesfully realized in the past (see e.g.\ Refs.~\onlinecite{tian2021dyocl,Tian_2016,Wang_2016,huang2017layer}). A reduction of the saturation magnetization, for example in diluted ferromagnets, would also allow for the parameters in larger, more readily available junction geometries to be found in the desired region. It can thus be said, that the predicted BKT transition likely resides at the boundary between current experimental capabilities and what are foreseeable advancements.

It is worth noting that generally a larger number of fermion flavors stabilizes the mean-field results, where an ordinary (complex) fermion can be understood as two (real) Majorana fermions. Therefore, if we considered a one-dimensional channel of ``ordinary'' instead of Majorana modes (e.g. quantum Hall edge states), this BKT transition would likely not be attainable in realistic systems, since in that case even smaller stiffnesses of the magnetization would be necessary to access it. It is the Majorana nature of the fermions that allows the transition to take place at higher values of $\tilde{M}$ and $\tilde{A}$.

In contrast, for the TCI transition at strong easy-axis anisotropies discussed in Sec.~\ref{sec:strong_easy-axis}, consider the case where $\tilde{B}\sim \tilde{g}\sim 1$. With the estimates from Ref.~\onlinecite{reich_magnetization_2023} (assuming a ratio $\tilde{A}\tilde{B}/\tilde{M}^2 \sim 10^7$ in accordance with the experimentally relevant values), we find for this transition critical stiffnesses of $\tilde{M}^2\tilde{A} \sim 10^5$. According to our estimates above, this transition is thus accessible by devices with a more readily attainable magnetic cross-sectional area of $DW \sim 100$\,nm$^2$.
}

\section{Conjectured phase structure in strong-coupling regime and full phase diagram \label{sec:conjecture}}

As mentioned above, within our framework it is not possible to rigorously differentiate between the different strong-coupling phases which may emerge for $B>0$. Let us here speculate about this part of the phase diagram, which is not accessible to our calculations, and subsequently present a conjecture for the full phase diagram, resulting from combining these speculations with our analysis in this paper. 

We expect three strong-coupling phases to exist for $B>0$:
\begin{itemize}
\item[(i)] one with the $x$-inversion symmetry spontaneously broken ($\braket{m_x} = \pm 1$, purple in Fig.~\ref{fig:full-phase-diagram}), 
\item[(ii)] one with $z$-inversion symmetry spontaneously broken ($\braket{m_z} = \pm 1$, blue in Fig.~\ref{fig:full-phase-diagram}),
\item[(iii)] and one with both $x$- and $z$-inversion broken ($0<|\braket{m_x}|,|\braket{m_z}|<1$, pink in Fig.~\ref{fig:full-phase-diagram}).
\end{itemize}

The latter two phases are the ones identified and discussed in Section \ref{sec:strong_easy-axis}, expected to be separated by a TCI transition. This claim can be further supported by noticing that the large $B$-limit corresponds to $\beta\Psi$ being restricted to small values in (\ref{eq:effective-action-psi}). Expanding the cosine terms for small arguments then yields the action of the Gross-Neveu-Yukawa model, for which a TCI transition has been theorized in Ref.~\onlinecite{grover_emergent_2014}.

In contrast, the transition between (i) and (iii) is one between two massive phases. It seems therefore most natural for it to belong to the Ising universality class with $c=1/2$. This is supported by the fact that close to the $x$-axis, where this transition happens, i.e. at $\eta \simeq \pi/2$ and $\psi \simeq 0$ or $\pi$, integrating out the (massive) Majorana fermions, one arrives at an effective $\phi^4$-theory. 

The way in which the TCI and the Ising critical line connect to the BKT transition remains an open question to be investigated. A relation to the considerations in Ref.~\onlinecite{huijse_emergent_2015}, where a {\color{black} supersymmetric} multicritical point at the intersection of an Ising and a BKT transition is identified, seems very likely. 
{\color{black} A possible connection to such kind of a supersymmetric multicriticality is further supported by the similarity between the supersymmetric sine-Gordon model and the effective theory that arises here.}

Together with the analysis and discussion presented in Sections \ref{sec:strong_easy-axis}, \ref{sec:easy-plane} and \ref{sec:easy-axis}, we arrive at the conjectured phase diagram presented in Fig.~\ref{fig:full-phase-diagram}.

\section{Summary and conclusions}

In this work, we propose topological superconductor-ferromagnet-superconductor junctions as a platform for hosting $1+1$-dimensional BKT and TCI phase transitions. Criticality arises from the coupling between counter-propagating Majorana fermions and a dynamic magnetization, which fluctuates either around an easy-axis or within an easy-plane.

In the easy-plane case, such a BKT transition separates a massless phase, with unbounded fluctuations of the magnetization within the easy-plane, from a massive phase with gapped out Majorana fermions and the magnetization pinned to the axis perpendicular to the junction, spontaneously breaking a $\mathbb{Z}_2$ symmetry. The massive phase is stabilized by the spatial and temporal rigidity of the magnetic modes. 

For the easy-axis anisotropy, the derivation of an effective theory is less straightforward than in the easy-plane case. Starting from mean-field considerations, we arrived at a conjecture for an effective low-energy theory based on an emerging effective easy-plane (oriented perpendicular to the propagation direction of the Majorana modes). Therein, we again identified a BKT transition separating a phase with unbounded fluctuations in the effective easy-plane from a strong-coupling regime with pinned magnetization, leading either to a phase with massive fermions or, if the magnetization is pinned to the easy-axis direction, a different phase with critical Majoranas. This pinning of the magnetization is again enabled by the magnetic stiffnesses dampening the fluctuations.

Finally, we presented a suggestion for the phase diagram of the topological SMS junction spanning all values of the magnetic anisotropy, based on mutually complementary mean-field and RG arguments. For the regions in parameter space not accessible within our framework, we provided speculative conjectures including a possible multicriticality.

\section*{Acknowledgments}
We are indebted to J. Schmalian and A. D. Mirlin for numerous illuminating and ideas generating discussions. AS is grateful to V. Yakovenko and J. Sau for very useful discussions. This work was supported by DFG via the grant SH 81/7-1. DSS acknowledges  funding by DFG grant MI~658/13-1 and partial support of the Helmholtz Validation Fund project ``Qruise".

\onecolumngrid

\appendix 

\section{Derivation of RG flow equations \label{app:rg}}

In the following we are going to sketch the derivation of the flow equations (\ref{eq:flow-c})-(\ref{eq:flow-g}) from the action (\ref{eq:effective-action}), where, for convenience, we shift $\beta\phi\rightarrow\beta\phi-\frac{\pi}{2}$
\begin{align}
    S = \int d\tau\,dy\,&\Big[\underbrace{\frac{v}{2}\bar\chi\slashed\partial\chi + \frac{1}{2}\left(\frac{1}{c}(\partial_\tau\phi)^2+c(\partial_x\phi)^2\right)}_{S_0} \underbrace{-u\cos 2\beta\phi + \frac{g}{2}\bar\chi\chi\sin\beta\phi}_{S_{\rm int}}\Big]. \label{eq:effective-action-in-appendix}
\end{align}
To this end, we employ a Wilson RG and choose a scheme in which the bosonic velocity $c$ flows while the fermionic one $v$ stays fixed. Following the standard approach\cite{gogolin_bosonization_2004, giamarchi_quantum_2004}, the fields are split into slow ($\omega^2/v^2+q^2<\Lambda^2/b^2$) and fast ($\Lambda^2/b^2<\omega^2/v^2+q^2<\Lambda^2$) components (with $v$ replaced by $c$ for the bosonic fields), $\chi = \chi_s+\chi_f$, $\phi = \phi_s+\phi_f$, and $b=e^l$, $l\ll 1$. For the partition function one then has
\begin{align}
    Z = \int\mathcal{D}s\,\mathcal{D}f\,e^{-S_0[s]-S_0[f]-S_{\rm int}[s+f]} = Z_0^f\int\mathcal{D}s\,e^{-S_0[s]}\braket{e^{-S_{\rm int}[s+f]}}_f,
\end{align}
where $\braket{\mathcal{O}}_f \equiv \frac{1}{Z_0^f}\int\mathcal{D}f\,\mathcal{O}\,e^{-S_0[f]}$ denotes averaging over the fast fields and $Z_0^f\equiv \int\mathcal{D}f\,e^{-S_0[f]}$. With this, an effective action only containing slow modes can be obtained via
\begin{align}
    S_{\rm eff}[s] = S_0[s] - \log\braket{e^{-S_{\rm int}[s+f]}}_f = S_0[s] + S^{(1)}[s] - \frac{1}{2}\left[S^{(2)}[s]-(S^{(1)}[s])^2\right] + \mathcal{O}(u^3,g^3,u^2g,ug^2) \label{eq:action-expansion-in-app}
\end{align}
where $S^{(n)}[s] \equiv \braket{(S_{\rm int}[s+f])^n}_f$. \\

The first order contributions evaluate to
\begin{align}
    \braket{\cos 2\beta\phi}_f = \cos 2\beta\phi_s\left(1-\frac{\beta^2}{\pi}l\right),\quad \braket{\bar\chi\chi\sin\beta\phi}_f = \bar\chi_s\chi_s\sin\beta\phi_s\left(1-\frac{\beta^2}{4\pi}l\right),
\end{align}
such that the flow equations at this level read
\begin{align}
    \frac{du}{dl} = 2u \left(1-\frac{\beta^2}{2\pi}\right),\quad \frac{dg}{dl} = g\left(1-\frac{\beta^2}{4\pi}\right).
\end{align}
Note that the mass and self-interaction of the bosonic field become irrelevant for $\beta^2>2\pi$, while the fermion-boson interaction becomes irrelevant for $\beta^2>4\pi$.\\

For the second order terms, one needs the fact that
\begin{align}
\begin{split}
    &\braket{\cos 2\beta\phi(r)\cos 2\beta\phi(r^\prime)}_f - \braket{\cos 2\beta\phi(r)}_f\braket{\cos 2\beta\phi(r^\prime)}_f \\ 
    &= \frac{1}{2}\sum_{\sigma=\pm}\cos 2\beta(\phi_s(r)+\sigma\phi_s(r^\prime))e^{-2\beta^2 l/\pi}\left[\exp\left(-4\sigma\beta^2\int_f\frac{d^2p}{(2\pi)^2}\frac{\cos p\cdot(r-r^\prime)}{p^2}\right)-1\right],
\end{split}
\end{align}
where $r=(y,c\tau)$, $p=(q,\omega/c)$ and $\int_fd^2p$ implies integration over the fast momenta only. Since the integral over $p$ only includes values $|p|\simeq\Lambda$, the main contribution to it stems from $|r-r^\prime|\lesssim \Lambda^{-1}$. Followingly, we can introduce the center of mass $R=(r+r^\prime)/2$ and relative coordinates $\delta r = r-r^\prime$ and expand in small $\Lambda\,\delta r$.

For $\sigma = +1$ then, the resulting operator is proportional to $\cos 4\beta\phi_s$, which we have seen to be irrelevant if $\beta^2>\pi/2$. Since we will focus on the region where either $u$ or $g$ are marginal, i.e. $2\pi\lesssim\beta^2\lesssim 4\pi$, we discard this term and only keep the $\sigma=-1$-contribution. 

Performing a gradient expansion of the cosine
\begin{align}
    \cos 2\beta(\phi_s(r)-\phi_s(r^\prime)) \simeq \text{const.} - 2\beta^2(\delta r\cdot\partial_R\phi_s)^2,
\end{align}
we finally find 
\begin{align}
    \int d^2r d^2r^\prime \left[\braket{\cos 2\beta\phi(r)\cos 2\beta\phi(r^\prime)}_f - \braket{\cos 2\beta\phi(r)}_f\braket{\cos 2\beta\phi(r^\prime)}_f\right] \simeq -\frac{2\beta^4 C_1}{\Lambda^4}l \int d^2R\,(\partial_R\phi_s)^2,
\end{align}
where we introduced the numerical constant $C_1 \equiv \int_0^\infty d\rho\,\rho^3F(\rho) >0$ with $\int_f\frac{d^2p}{(2\pi)^2}\frac{\cos p\cdot r}{p^2} \equiv \frac{l}{2\pi}F(\Lambda|r|)+\mathcal{O}(l^2)$. For the plots in the main text we used $C_1=8$, as follows from employing a Gaussian cut-off $\int_0^\Lambda dp \rightarrow \int_0^\infty dp\,e^{-p^2/\Lambda^2}$.\\

We proceed similarly for the remaining second-order terms, discarding the generated 4-Majorana- and $\bar\chi\chi\sin 3\beta\phi$-terms as irrelevant in the region of interest. We find 
\begin{align}
\begin{split}
    \int d^2 r\,d^2r^\prime\,&\braket{\bar\chi_f(r)\chi_f(r)\bar\chi_f(r^\prime)\chi_f(r^\prime)\sin\beta\phi(r)\sin\beta\phi(r^\prime)}_f \\ 
    &\simeq -\frac{lc}{\pi v} \int d^2R\,\cos2\beta\phi_s(R) - \frac{l\beta^2 c^2}{2\pi v^2\Lambda^2}\int d\tau\,dy\,\left[\frac{1}{v}(\partial_\tau\phi_s)^2+v(\partial_y\phi_s)^2\right]
\end{split}
\end{align}
as well as
\begin{align}
\begin{split}
    \int d^2 r\,d^2r^\prime\,\bar\chi_s(r)\chi_s(r)&\left[\braket{\sin\beta\phi(r)\cos 2\beta\phi(r^\prime)}_f-\braket{\sin\beta\phi(r)}_f\braket{\cos 2\beta\phi(r^\prime)}_f\right] \\
    &\simeq -\frac{\beta^2C_2}{\Lambda^2}l\int d^2R\,\bar\chi_s(R)\chi_s(R)\sin\beta\phi_s(R)
\end{split}
\end{align}
with the numerical constant $C_2 \equiv \int_0^\infty d\rho\,\rho F(\rho) >0$ and $C_2=2$ for the Gaussian cut-off. \\

Plugging these results into (\ref{eq:action-expansion-in-app}) yields the flow equations (\ref{eq:flow-c})-(\ref{eq:flow-g}) given in the main text.

\section{RG flow taking into account $A^\prime\Lambda^2 \gg (-B)$ \label{app:checkAprime}}

Even though the parameter $A^\prime$ in Eq.~(\ref{eq:action-with-Aprime}) is strongly irrelevant, in the case $A^\prime\Lambda^2\gg (-B)$ it is not immediately clear how the (at least initially not neglectable) influence of $A^\prime$ affects the flow. Let us therefore rederive the RG equations at lowest order in $\tilde{u}$ and $\tilde{g}$ taking $A^\prime$ into account. We can write the relevant action as 
\begin{align}
    S_{A^\prime} = S + \frac{1}{2}\int \frac{dq\,d\omega}{(2\pi)^2} \frac{\omega^2}{c_0} \frac{A^\prime q^2/B}{1-A^\prime q^2/B}|\phi_{q,\omega}|^2,
\end{align}
where $S$ is the action given in Eq.~(\ref{eq:effective-action}) and $c_0=c(l=0)$ will not be renormalized. With that, one finds 
\begin{gather}
    \frac{d\tilde{A}^\prime}{dl} = -2\tilde{A}^\prime, \qquad \frac{d\tilde{u}}{dl} = 2\tilde{u}\left(1-\Gamma(\tilde{A}^\prime)\frac{\beta^2}{2\pi}\right), \qquad
    \frac{d\tilde{g}}{dl} = \tilde{g}\left(1-\Gamma(\tilde{A}^\prime)\frac{\beta^2}{4\pi}\right) \label{eq:flow-with-Aprime}
\end{gather}
with
\begin{align}
    \Gamma(\tilde{A}^\prime) = \frac{1}{2\pi}\int_0^{2\pi}d\theta\,\frac{1+\tilde{A^\prime}\cos^2\theta}{\left(\frac{v}{c}\sin^2\theta + \frac{c}{v}
    \cos^2\theta\right)\left(1+\tilde{A}^\prime\cos^2\theta\right)-\frac{v\tilde{A}^\prime}{c_0}\sin^2\theta\cos^2\theta},
\end{align}
where $\tilde{A}^\prime = A^\prime\Lambda^2/(-B)$. Note that $\Gamma(0) = 1$. Let us here for simplicity consider the Lorentz symmetric situation $v=c=c_0$. Then 
\begin{align}
    \Gamma(\tilde{A}^\prime) = \left(1+\tilde{A}^\prime\right)^{1/4}\cos\left(\frac{1}{2}\arctan\sqrt{\tilde{A}^\prime}\right).
\end{align}

\begin{figure}
    \centering
    \includegraphics[width=0.5\textwidth]{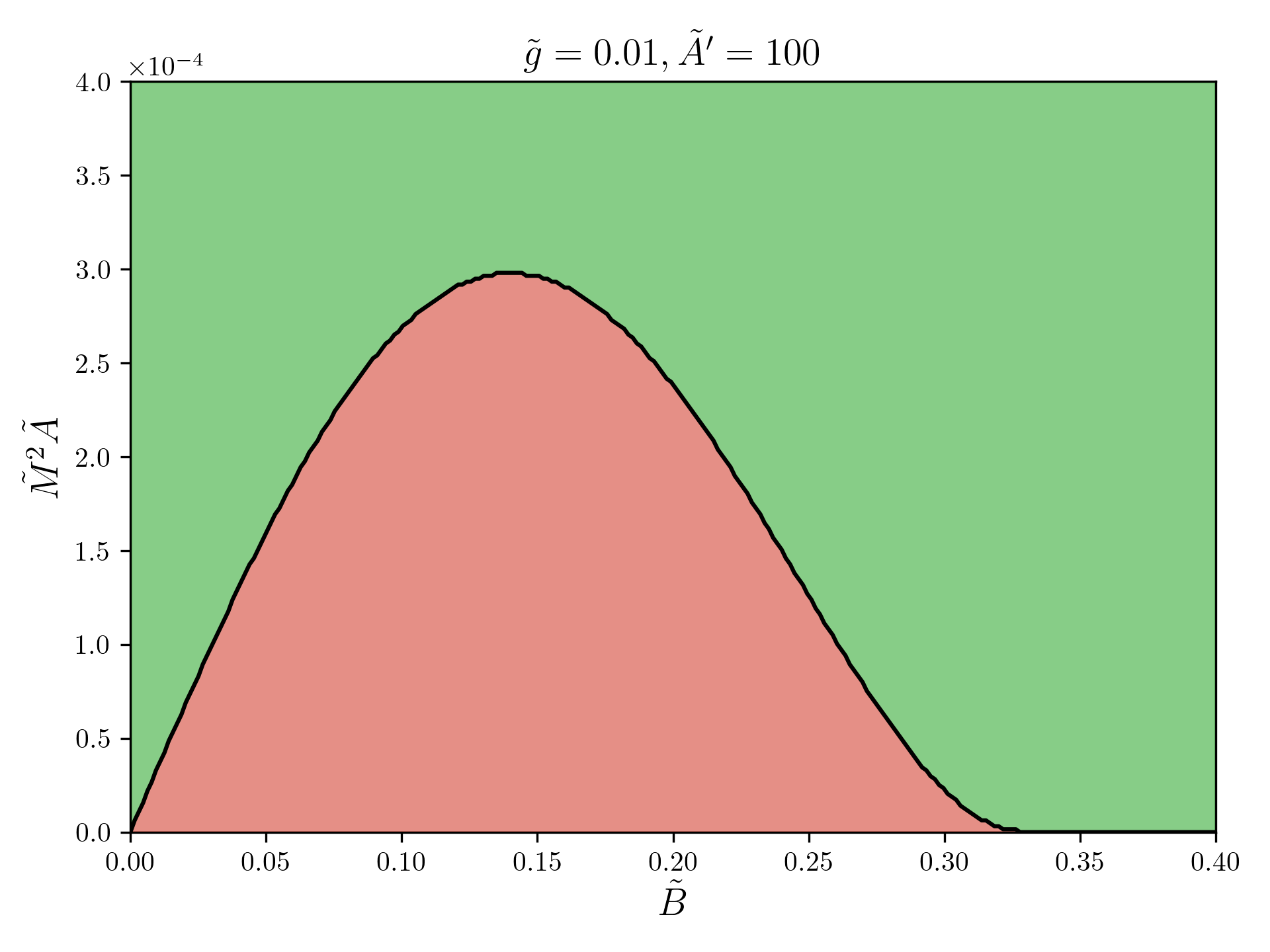}
    \caption{Same phase diagram as in Fig.~\ref{fig:phasediagram_easy-axis_g=0.01}, but with $\tilde{A}^\prime=100$ taken into account. Note the different scale compared to Fig.~\ref{fig:phasediagram_easy-axis_g=0.01}.}
    \label{fig:phasediagram_easy-axis_inclAprime}
\end{figure}

Generating in Fig.~\ref{fig:phasediagram_easy-axis_inclAprime} as an example the same phase diagram as shown in Fig.~\ref{fig:phasediagram_easy-axis_g=0.01} of the main text while taking $\Gamma(\tilde{A}^\prime)$ into account, with a value of $\tilde{A}^\prime(0)$ as large as $\tilde{A}^\prime(0)=100$, clearly shows that quantitative, but no qualitative corrections occur. Considering $A^\prime$ in second order in $\tilde{u}$ and $\tilde{g}$ makes the constants $C_{1/2}$ become dependent on $A^\prime$. We do not expect this either to lead to any qualitative changes to our findings.

\twocolumngrid

\bibliography{references,references_DS,references_experiment}

\begin{thebibliography}{39}%
\makeatletter
\providecommand \@ifxundefined [1]{%
 \@ifx{#1\undefined}
}%
\providecommand \@ifnum [1]{%
 \ifnum #1\expandafter \@firstoftwo
 \else \expandafter \@secondoftwo
 \fi
}%
\providecommand \@ifx [1]{%
 \ifx #1\expandafter \@firstoftwo
 \else \expandafter \@secondoftwo
 \fi
}%
\providecommand \natexlab [1]{#1}%
\providecommand \enquote  [1]{``#1''}%
\providecommand \bibnamefont  [1]{#1}%
\providecommand \bibfnamefont [1]{#1}%
\providecommand \citenamefont [1]{#1}%
\providecommand \href@noop [0]{\@secondoftwo}%
\providecommand \href [0]{\begingroup \@sanitize@url \@href}%
\providecommand \@href[1]{\@@startlink{#1}\@@href}%
\providecommand \@@href[1]{\endgroup#1\@@endlink}%
\providecommand \@sanitize@url [0]{\catcode `\\12\catcode `\$12\catcode `\&12\catcode `\#12\catcode `\^12\catcode `\_12\catcode `\%12\relax}%
\providecommand \@@startlink[1]{}%
\providecommand \@@endlink[0]{}%
\providecommand \url  [0]{\begingroup\@sanitize@url \@url }%
\providecommand \@url [1]{\endgroup\@href {#1}{\urlprefix }}%
\providecommand \urlprefix  [0]{URL }%
\providecommand \Eprint [0]{\href }%
\providecommand \doibase [0]{http://dx.doi.org/}%
\providecommand \selectlanguage [0]{\@gobble}%
\providecommand \bibinfo  [0]{\@secondoftwo}%
\providecommand \bibfield  [0]{\@secondoftwo}%
\providecommand \translation [1]{[#1]}%
\providecommand \BibitemOpen [0]{}%
\providecommand \bibitemStop [0]{}%
\providecommand \bibitemNoStop [0]{.\EOS\space}%
\providecommand \EOS [0]{\spacefactor3000\relax}%
\providecommand \BibitemShut  [1]{\csname bibitem#1\endcsname}%
\let\auto@bib@innerbib\@empty
\bibitem [{\citenamefont {Qi}\ and\ \citenamefont {Zhang}(2011)}]{RevModPhys.83.1057}%
  \BibitemOpen
  \bibfield  {author} {\bibinfo {author} {\bibfnamefont {Xiao-Liang}\ \bibnamefont {Qi}}\ and\ \bibinfo {author} {\bibfnamefont {Shou-Cheng}\ \bibnamefont {Zhang}},\ }\bibfield  {title} {\enquote {\bibinfo {title} {Topological insulators and superconductors},}\ }\href {\doibase 10.1103/RevModPhys.83.1057} {\bibfield  {journal} {\bibinfo  {journal} {Rev. Mod. Phys.}\ }\textbf {\bibinfo {volume} {83}},\ \bibinfo {pages} {1057--1110} (\bibinfo {year} {2011})}\BibitemShut {NoStop}%
\bibitem [{\citenamefont {Alicea}(2012)}]{alicea_new_2012}%
  \BibitemOpen
  \bibfield  {author} {\bibinfo {author} {\bibfnamefont {Jason}\ \bibnamefont {Alicea}},\ }\bibfield  {title} {\enquote {\bibinfo {title} {New directions in the pursuit of {Majorana} fermions in solid state systems},}\ }\href {\doibase 10.1088/0034-4885/75/7/076501} {\bibfield  {journal} {\bibinfo  {journal} {Reports on Progress in Physics}\ }\textbf {\bibinfo {volume} {75}},\ \bibinfo {pages} {076501} (\bibinfo {year} {2012})}\BibitemShut {NoStop}%
\bibitem [{\citenamefont {Beenakker}(2013)}]{Beenakker-Rev}%
  \BibitemOpen
  \bibfield  {author} {\bibinfo {author} {\bibfnamefont {C.W.J.}\ \bibnamefont {Beenakker}},\ }\bibfield  {title} {\enquote {\bibinfo {title} {Search for majorana fermions in superconductors},}\ }\href {\doibase 10.1146/annurev-conmatphys-030212-184337} {\bibfield  {journal} {\bibinfo  {journal} {Annual Review of Condensed Matter Physics}\ }\textbf {\bibinfo {volume} {4}},\ \bibinfo {pages} {113--136} (\bibinfo {year} {2013})}\BibitemShut {NoStop}%
\bibitem [{\citenamefont {M{\'e}nard}\ \emph {et~al.}(2017)\citenamefont {M{\'e}nard}, \citenamefont {Guissart}, \citenamefont {Brun}, \citenamefont {Leriche}, \citenamefont {Trif}, \citenamefont {Debontridder}, \citenamefont {Demaille}, \citenamefont {Roditchev}, \citenamefont {Simon},\ and\ \citenamefont {Cren}}]{menard2017two}%
  \BibitemOpen
  \bibfield  {author} {\bibinfo {author} {\bibfnamefont {Gerbold~C}\ \bibnamefont {M{\'e}nard}}, \bibinfo {author} {\bibfnamefont {S{\'e}bastien}\ \bibnamefont {Guissart}}, \bibinfo {author} {\bibfnamefont {Christophe}\ \bibnamefont {Brun}}, \bibinfo {author} {\bibfnamefont {Rapha{\"e}l~T}\ \bibnamefont {Leriche}}, \bibinfo {author} {\bibfnamefont {Mircea}\ \bibnamefont {Trif}}, \bibinfo {author} {\bibfnamefont {Fran{\c{c}}ois}\ \bibnamefont {Debontridder}}, \bibinfo {author} {\bibfnamefont {Dominique}\ \bibnamefont {Demaille}}, \bibinfo {author} {\bibfnamefont {Dimitri}\ \bibnamefont {Roditchev}}, \bibinfo {author} {\bibfnamefont {Pascal}\ \bibnamefont {Simon}}, \ and\ \bibinfo {author} {\bibfnamefont {Tristan}\ \bibnamefont {Cren}},\ }\bibfield  {title} {\enquote {\bibinfo {title} {Two-dimensional topological superconductivity in pb/co/si (111)},}\ }\href {https://doi.org/10.1038/s41467-017-02192-x} {\bibfield  {journal} {\bibinfo  {journal} {Nature communications}\ }\textbf {\bibinfo {volume} {8}},\
  \bibinfo {pages} {2040} (\bibinfo {year} {2017})}\BibitemShut {NoStop}%
\bibitem [{\citenamefont {Palacio-Morales}\ \emph {et~al.}(2019)\citenamefont {Palacio-Morales}, \citenamefont {Mascot}, \citenamefont {Cocklin}, \citenamefont {Kim}, \citenamefont {Rachel}, \citenamefont {Morr},\ and\ \citenamefont {Wiesendanger}}]{doi:10.1126/sciadv.aav6600}%
  \BibitemOpen
  \bibfield  {author} {\bibinfo {author} {\bibfnamefont {Alexandra}\ \bibnamefont {Palacio-Morales}}, \bibinfo {author} {\bibfnamefont {Eric}\ \bibnamefont {Mascot}}, \bibinfo {author} {\bibfnamefont {Sagen}\ \bibnamefont {Cocklin}}, \bibinfo {author} {\bibfnamefont {Howon}\ \bibnamefont {Kim}}, \bibinfo {author} {\bibfnamefont {Stephan}\ \bibnamefont {Rachel}}, \bibinfo {author} {\bibfnamefont {Dirk~K.}\ \bibnamefont {Morr}}, \ and\ \bibinfo {author} {\bibfnamefont {Roland}\ \bibnamefont {Wiesendanger}},\ }\bibfield  {title} {\enquote {\bibinfo {title} {Atomic-scale interface engineering of majorana edge modes in a 2d magnet-superconductor hybrid system},}\ }\href {\doibase 10.1126/sciadv.aav6600} {\bibfield  {journal} {\bibinfo  {journal} {Science Advances}\ }\textbf {\bibinfo {volume} {5}},\ \bibinfo {pages} {eaav6600} (\bibinfo {year} {2019})}\BibitemShut {NoStop}%
\bibitem [{\citenamefont {He}\ \emph {et~al.}(2017)\citenamefont {He}, \citenamefont {Pan}, \citenamefont {Stern}, \citenamefont {Burks}, \citenamefont {Che} \emph {et~al.}}]{HeScience2017}%
  \BibitemOpen
  \bibfield  {author} {\bibinfo {author} {\bibfnamefont {Qing~Lin}\ \bibnamefont {He}}, \bibinfo {author} {\bibfnamefont {Lei}\ \bibnamefont {Pan}}, \bibinfo {author} {\bibfnamefont {Alexander~L.}\ \bibnamefont {Stern}}, \bibinfo {author} {\bibfnamefont {Edward~C.}\ \bibnamefont {Burks}}, \bibinfo {author} {\bibfnamefont {Xiaoyu}\ \bibnamefont {Che}},  \emph {et~al.},\ }\bibfield  {title} {\enquote {\bibinfo {title} {Chiral majorana fermion modes in a quantum anomalous hall insulator{\textendash}superconductor structure},}\ }\href {\doibase 10.1126/science.aag2792} {\bibfield  {journal} {\bibinfo  {journal} {Science}\ }\textbf {\bibinfo {volume} {357}},\ \bibinfo {pages} {294--299} (\bibinfo {year} {2017})}\BibitemShut {NoStop}%
\bibitem [{\citenamefont {Shen}\ \emph {et~al.}(2020)\citenamefont {Shen}, \citenamefont {Lyu}, \citenamefont {Gao}, \citenamefont {Xie}, \citenamefont {Chen}, \citenamefont {Cho}, \citenamefont {Atanov}, \citenamefont {Chen}, \citenamefont {Liu}, \citenamefont {Hu}, \citenamefont {Yip}, \citenamefont {Goh}, \citenamefont {He}, \citenamefont {Pan}, \citenamefont {Wang}, \citenamefont {Law},\ and\ \citenamefont {Lortz}}]{shen2018spectroscopic}%
  \BibitemOpen
  \bibfield  {author} {\bibinfo {author} {\bibfnamefont {Junying}\ \bibnamefont {Shen}}, \bibinfo {author} {\bibfnamefont {Jian}\ \bibnamefont {Lyu}}, \bibinfo {author} {\bibfnamefont {Jason~Z.}\ \bibnamefont {Gao}}, \bibinfo {author} {\bibfnamefont {Ying-Ming}\ \bibnamefont {Xie}}, \bibinfo {author} {\bibfnamefont {Chui-Zhen}\ \bibnamefont {Chen}}, \bibinfo {author} {\bibfnamefont {Chang-woo}\ \bibnamefont {Cho}}, \bibinfo {author} {\bibfnamefont {Omargeldi}\ \bibnamefont {Atanov}}, \bibinfo {author} {\bibfnamefont {Zhijie}\ \bibnamefont {Chen}}, \bibinfo {author} {\bibfnamefont {Kai}\ \bibnamefont {Liu}}, \bibinfo {author} {\bibfnamefont {Yajian~J.}\ \bibnamefont {Hu}}, \bibinfo {author} {\bibfnamefont {King~Yau}\ \bibnamefont {Yip}}, \bibinfo {author} {\bibfnamefont {Swee~K.}\ \bibnamefont {Goh}}, \bibinfo {author} {\bibfnamefont {Qing~Lin}\ \bibnamefont {He}}, \bibinfo {author} {\bibfnamefont {Lei}\ \bibnamefont {Pan}}, \bibinfo {author} {\bibfnamefont {Kang~L.}\ \bibnamefont {Wang}}, \bibinfo {author}
  {\bibfnamefont {Kam~Tuen}\ \bibnamefont {Law}}, \ and\ \bibinfo {author} {\bibfnamefont {Rolf}\ \bibnamefont {Lortz}},\ }\bibfield  {title} {\enquote {\bibinfo {title} {Spectroscopic fingerprint of chiral majorana modes at the edge of a quantum anomalous hall insulator/superconductor heterostructure},}\ }\href {\doibase 10.1073/pnas.1910967117} {\bibfield  {journal} {\bibinfo  {journal} {Proceedings of the National Academy of Sciences}\ }\textbf {\bibinfo {volume} {117}},\ \bibinfo {pages} {238--242} (\bibinfo {year} {2020})}\BibitemShut {NoStop}%
\bibitem [{\citenamefont {Kasahara}\ \emph {et~al.}(2018)\citenamefont {Kasahara}, \citenamefont {Ohnishi}, \citenamefont {Mizukami}, \citenamefont {Tanaka}, \citenamefont {Ma}, \citenamefont {Sugii}, \citenamefont {Kurita}, \citenamefont {Tanaka}, \citenamefont {Nasu}, \citenamefont {Motome}, \citenamefont {Shibauchi},\ and\ \citenamefont {Matsuda}}]{kasahara2018majorana}%
  \BibitemOpen
  \bibfield  {author} {\bibinfo {author} {\bibfnamefont {Y.}~\bibnamefont {Kasahara}}, \bibinfo {author} {\bibfnamefont {T.}~\bibnamefont {Ohnishi}}, \bibinfo {author} {\bibfnamefont {Y.}~\bibnamefont {Mizukami}}, \bibinfo {author} {\bibfnamefont {O.}~\bibnamefont {Tanaka}}, \bibinfo {author} {\bibfnamefont {Sixiao}\ \bibnamefont {Ma}}, \bibinfo {author} {\bibfnamefont {K.}~\bibnamefont {Sugii}}, \bibinfo {author} {\bibfnamefont {N.}~\bibnamefont {Kurita}}, \bibinfo {author} {\bibfnamefont {H.}~\bibnamefont {Tanaka}}, \bibinfo {author} {\bibfnamefont {J.}~\bibnamefont {Nasu}}, \bibinfo {author} {\bibfnamefont {Y.}~\bibnamefont {Motome}}, \bibinfo {author} {\bibfnamefont {T.}~\bibnamefont {Shibauchi}}, \ and\ \bibinfo {author} {\bibfnamefont {Y.}~\bibnamefont {Matsuda}},\ }\bibfield  {title} {\enquote {\bibinfo {title} {Majorana quantization and half-integer thermal quantum hall effect in a kitaev spin liquid},}\ }\href {\doibase 10.1038/s41586-018-0274-0} {\bibfield  {journal} {\bibinfo  {journal} {Nature}\
  }\textbf {\bibinfo {volume} {559}},\ \bibinfo {pages} {227--231} (\bibinfo {year} {2018})}\BibitemShut {NoStop}%
\bibitem [{\citenamefont {Wang}\ \emph {et~al.}(2020)\citenamefont {Wang}, \citenamefont {Rodriguez}, \citenamefont {Jiao}, \citenamefont {Howard}, \citenamefont {Graham}, \citenamefont {Gu}, \citenamefont {Hughes}, \citenamefont {Morr},\ and\ \citenamefont {Madhavan}}]{Wang104}%
  \BibitemOpen
  \bibfield  {author} {\bibinfo {author} {\bibfnamefont {Zhenyu}\ \bibnamefont {Wang}}, \bibinfo {author} {\bibfnamefont {Jorge~Olivares}\ \bibnamefont {Rodriguez}}, \bibinfo {author} {\bibfnamefont {Lin}\ \bibnamefont {Jiao}}, \bibinfo {author} {\bibfnamefont {Sean}\ \bibnamefont {Howard}}, \bibinfo {author} {\bibfnamefont {Martin}\ \bibnamefont {Graham}}, \bibinfo {author} {\bibfnamefont {G.~D.}\ \bibnamefont {Gu}}, \bibinfo {author} {\bibfnamefont {Taylor~L.}\ \bibnamefont {Hughes}}, \bibinfo {author} {\bibfnamefont {Dirk~K.}\ \bibnamefont {Morr}}, \ and\ \bibinfo {author} {\bibfnamefont {Vidya}\ \bibnamefont {Madhavan}},\ }\bibfield  {title} {\enquote {\bibinfo {title} {Evidence for dispersing 1d majorana channels in an iron-based superconductor},}\ }\href {\doibase 10.1126/science.aaw8419} {\bibfield  {journal} {\bibinfo  {journal} {Science}\ }\textbf {\bibinfo {volume} {367}},\ \bibinfo {pages} {104--108} (\bibinfo {year} {2020})}\BibitemShut {NoStop}%
\bibitem [{\citenamefont {Kezilebieke}\ \emph {et~al.}(2020)\citenamefont {Kezilebieke}, \citenamefont {Huda}, \citenamefont {Va{\v n}o}, \citenamefont {Aapro}, \citenamefont {Ganguli}, \citenamefont {Silveira}, \citenamefont {G{\l}odzik}, \citenamefont {Foster}, \citenamefont {Ojanen},\ and\ \citenamefont {Liljeroth}}]{Kezilebieke:2020ab}%
  \BibitemOpen
  \bibfield  {author} {\bibinfo {author} {\bibfnamefont {Shawulienu}\ \bibnamefont {Kezilebieke}}, \bibinfo {author} {\bibfnamefont {Md~Nurul}\ \bibnamefont {Huda}}, \bibinfo {author} {\bibfnamefont {Viliam}\ \bibnamefont {Va{\v n}o}}, \bibinfo {author} {\bibfnamefont {Markus}\ \bibnamefont {Aapro}}, \bibinfo {author} {\bibfnamefont {Somesh~C.}\ \bibnamefont {Ganguli}}, \bibinfo {author} {\bibfnamefont {Orlando~J.}\ \bibnamefont {Silveira}}, \bibinfo {author} {\bibfnamefont {Szczepan}\ \bibnamefont {G{\l}odzik}}, \bibinfo {author} {\bibfnamefont {Adam~S.}\ \bibnamefont {Foster}}, \bibinfo {author} {\bibfnamefont {Teemu}\ \bibnamefont {Ojanen}}, \ and\ \bibinfo {author} {\bibfnamefont {Peter}\ \bibnamefont {Liljeroth}},\ }\bibfield  {title} {\enquote {\bibinfo {title} {Topological superconductivity in a van der waals heterostructure},}\ }\href {\doibase 10.1038/s41586-020-2989-y} {\bibfield  {journal} {\bibinfo  {journal} {Nature}\ }\textbf {\bibinfo {volume} {588}},\ \bibinfo {pages} {424--428} (\bibinfo
  {year} {2020})}\BibitemShut {NoStop}%
\bibitem [{\citenamefont {Li}\ \emph {et~al.}(2024)\citenamefont {Li}, \citenamefont {Yin}, \citenamefont {Li}, \citenamefont {Gong}, \citenamefont {Chen}, \citenamefont {Zhang}, \citenamefont {Yan},\ and\ \citenamefont {Feng}}]{Li2024}%
  \BibitemOpen
  \bibfield  {author} {\bibinfo {author} {\bibfnamefont {Yuanji}\ \bibnamefont {Li}}, \bibinfo {author} {\bibfnamefont {Ruotong}\ \bibnamefont {Yin}}, \bibinfo {author} {\bibfnamefont {Mingzhe}\ \bibnamefont {Li}}, \bibinfo {author} {\bibfnamefont {Jiashuo}\ \bibnamefont {Gong}}, \bibinfo {author} {\bibfnamefont {Ziyuan}\ \bibnamefont {Chen}}, \bibinfo {author} {\bibfnamefont {Jiakang}\ \bibnamefont {Zhang}}, \bibinfo {author} {\bibfnamefont {Ya-Jun}\ \bibnamefont {Yan}}, \ and\ \bibinfo {author} {\bibfnamefont {Dong-Lai}\ \bibnamefont {Feng}},\ }\bibfield  {title} {\enquote {\bibinfo {title} {Observation of {Y}u-{S}hiba-{R}usinov-like states at the edge of {C}r{B}r3/{N}b{S}e2 heterostructure},}\ }\href {\doibase 10.1038/s41467-024-54525-2} {\bibfield  {journal} {\bibinfo  {journal} {Nature Communications}\ }\textbf {\bibinfo {volume} {15}},\ \bibinfo {pages} {10121} (\bibinfo {year} {2024})}\BibitemShut {NoStop}%
\bibitem [{\citenamefont {Fu}\ and\ \citenamefont {Kane}(2008)}]{fu_superconducting_2008}%
  \BibitemOpen
  \bibfield  {author} {\bibinfo {author} {\bibfnamefont {Liang}\ \bibnamefont {Fu}}\ and\ \bibinfo {author} {\bibfnamefont {C~L}\ \bibnamefont {Kane}},\ }\bibfield  {title} {\enquote {\bibinfo {title} {Superconducting {Proximity} {Effect} and {Majorana} {Fermions} at the {Surface} of a {Topological} {Insulator}},}\ }\href {\doibase 10.1103/PhysRevLett.100.096407} {\bibfield  {journal} {\bibinfo  {journal} {Physical Review Letters}\ }\textbf {\bibinfo {volume} {100}},\ \bibinfo {pages} {096407} (\bibinfo {year} {2008})}\BibitemShut {NoStop}%
\bibitem [{\citenamefont {Fu}\ and\ \citenamefont {Kane}(2009)}]{FuKanePrl2009}%
  \BibitemOpen
  \bibfield  {author} {\bibinfo {author} {\bibfnamefont {Liang}\ \bibnamefont {Fu}}\ and\ \bibinfo {author} {\bibfnamefont {C.~L.}\ \bibnamefont {Kane}},\ }\bibfield  {title} {\enquote {\bibinfo {title} {Probing neutral majorana fermion edge modes with charge transport},}\ }\href {\doibase 10.1103/PhysRevLett.102.216403} {\bibfield  {journal} {\bibinfo  {journal} {Phys. Rev. Lett.}\ }\textbf {\bibinfo {volume} {102}},\ \bibinfo {pages} {216403} (\bibinfo {year} {2009})}\BibitemShut {NoStop}%
\bibitem [{\citenamefont {Tanaka}\ \emph {et~al.}(2009)\citenamefont {Tanaka}, \citenamefont {Yokoyama},\ and\ \citenamefont {Nagaosa}}]{tanaka_manipulation_2009}%
  \BibitemOpen
  \bibfield  {author} {\bibinfo {author} {\bibfnamefont {Yukio}\ \bibnamefont {Tanaka}}, \bibinfo {author} {\bibfnamefont {Takehito}\ \bibnamefont {Yokoyama}}, \ and\ \bibinfo {author} {\bibfnamefont {Naoto}\ \bibnamefont {Nagaosa}},\ }\bibfield  {title} {{\selectlanguage {en}\enquote {\bibinfo {title} {Manipulation of the {Majorana} {Fermion}, {Andreev} {Reflection}, and {Josephson} {Current} on {Topological} {Insulators}},}\ }}\href {\doibase 10.1103/PhysRevLett.103.107002} {\bibfield  {journal} {\bibinfo  {journal} {Physical Review Letters}\ }\textbf {\bibinfo {volume} {103}},\ \bibinfo {pages} {107002} (\bibinfo {year} {2009})}\BibitemShut {NoStop}%
\bibitem [{\citenamefont {Zyuzin}\ \emph {et~al.}(2016)\citenamefont {Zyuzin}, \citenamefont {Alidoust},\ and\ \citenamefont {Loss}}]{zyuzin_josephson_2016}%
  \BibitemOpen
  \bibfield  {author} {\bibinfo {author} {\bibfnamefont {Alexander}\ \bibnamefont {Zyuzin}}, \bibinfo {author} {\bibfnamefont {Mohammad}\ \bibnamefont {Alidoust}}, \ and\ \bibinfo {author} {\bibfnamefont {Daniel}\ \bibnamefont {Loss}},\ }\bibfield  {title} {{\selectlanguage {en}\enquote {\bibinfo {title} {Josephson {Junction} through a {Disordered} {Topological} {Insulator} with {Helical} {Magnetization}},}\ }}\href {\doibase 10.1103/PhysRevB.93.214502} {\bibfield  {journal} {\bibinfo  {journal} {Physical Review B}\ }\textbf {\bibinfo {volume} {93}},\ \bibinfo {pages} {214502} (\bibinfo {year} {2016})}\BibitemShut {NoStop}%
\bibitem [{\citenamefont {Bobkova}\ \emph {et~al.}(2016)\citenamefont {Bobkova}, \citenamefont {Bobkov}, \citenamefont {Zyuzin},\ and\ \citenamefont {Alidoust}}]{bobkova_magnetoelectrics_2016}%
  \BibitemOpen
  \bibfield  {author} {\bibinfo {author} {\bibfnamefont {I.~V.}\ \bibnamefont {Bobkova}}, \bibinfo {author} {\bibfnamefont {A.~M.}\ \bibnamefont {Bobkov}}, \bibinfo {author} {\bibfnamefont {Alexander~A.}\ \bibnamefont {Zyuzin}}, \ and\ \bibinfo {author} {\bibfnamefont {Mohammad}\ \bibnamefont {Alidoust}},\ }\bibfield  {title} {{\selectlanguage {en}\enquote {\bibinfo {title} {Magnetoelectrics in {Disordered} {Topological} {Insulator} {Josephson} {Junctions}},}\ }}\href {\doibase 10.1103/PhysRevB.94.134506} {\bibfield  {journal} {\bibinfo  {journal} {Physical Review B}\ }\textbf {\bibinfo {volume} {94}},\ \bibinfo {pages} {134506} (\bibinfo {year} {2016})}\BibitemShut {NoStop}%
\bibitem [{\citenamefont {Amundsen}\ \emph {et~al.}(2018)\citenamefont {Amundsen}, \citenamefont {Hugdal}, \citenamefont {Sudbo},\ and\ \citenamefont {Linder}}]{amundsen_vortex_2018}%
  \BibitemOpen
  \bibfield  {author} {\bibinfo {author} {\bibfnamefont {Morten}\ \bibnamefont {Amundsen}}, \bibinfo {author} {\bibfnamefont {Henning~G.}\ \bibnamefont {Hugdal}}, \bibinfo {author} {\bibfnamefont {Asle}\ \bibnamefont {Sudbo}}, \ and\ \bibinfo {author} {\bibfnamefont {Jacob}\ \bibnamefont {Linder}},\ }\bibfield  {title} {{\selectlanguage {en}\enquote {\bibinfo {title} {Vortex spin valve on a topological insulator},}\ }}\href {\doibase 10.1103/PhysRevB.98.144505} {\bibfield  {journal} {\bibinfo  {journal} {Physical Review B}\ }\textbf {\bibinfo {volume} {98}},\ \bibinfo {pages} {144505} (\bibinfo {year} {2018})}\BibitemShut {NoStop}%
\bibitem [{\citenamefont {Reich}\ \emph {et~al.}(2023)\citenamefont {Reich}, \citenamefont {Berg}, \citenamefont {Schmalian},\ and\ \citenamefont {Shnirman}}]{reich_magnetization_2023}%
  \BibitemOpen
  \bibfield  {author} {\bibinfo {author} {\bibfnamefont {Adrian}\ \bibnamefont {Reich}}, \bibinfo {author} {\bibfnamefont {Erez}\ \bibnamefont {Berg}}, \bibinfo {author} {\bibfnamefont {Jörg}\ \bibnamefont {Schmalian}}, \ and\ \bibinfo {author} {\bibfnamefont {Alexander}\ \bibnamefont {Shnirman}},\ }\bibfield  {title} {{\selectlanguage {en}\enquote {\bibinfo {title} {Magnetization dynamics and {Peierls} instability in topological {Josephson} structures},}\ }}\href {\doibase 10.1103/PhysRevB.107.245411} {\bibfield  {journal} {\bibinfo  {journal} {Physical Review B}\ }\textbf {\bibinfo {volume} {107}},\ \bibinfo {pages} {245411} (\bibinfo {year} {2023})}\BibitemShut {NoStop}%
\bibitem [{\citenamefont {Grover}\ \emph {et~al.}(2014)\citenamefont {Grover}, \citenamefont {Sheng},\ and\ \citenamefont {Vishwanath}}]{grover_emergent_2014}%
  \BibitemOpen
  \bibfield  {author} {\bibinfo {author} {\bibfnamefont {Tarun}\ \bibnamefont {Grover}}, \bibinfo {author} {\bibfnamefont {D.~N.}\ \bibnamefont {Sheng}}, \ and\ \bibinfo {author} {\bibfnamefont {Ashvin}\ \bibnamefont {Vishwanath}},\ }\bibfield  {title} {{\selectlanguage {en}\enquote {\bibinfo {title} {Emergent {Space}-{Time} {Supersymmetry} at the {Boundary} of a {Topological} {Phase}},}\ }}\href {\doibase 10.1126/science.1248253} {\bibfield  {journal} {\bibinfo  {journal} {Science}\ }\textbf {\bibinfo {volume} {344}},\ \bibinfo {pages} {280--283} (\bibinfo {year} {2014})}\BibitemShut {NoStop}%
\bibitem [{\citenamefont {Rahmani}\ \emph {et~al.}(2015)\citenamefont {Rahmani}, \citenamefont {Zhu}, \citenamefont {Franz},\ and\ \citenamefont {Affleck}}]{rahmani_emergent_2015}%
  \BibitemOpen
  \bibfield  {author} {\bibinfo {author} {\bibfnamefont {Armin}\ \bibnamefont {Rahmani}}, \bibinfo {author} {\bibfnamefont {Xiaoyu}\ \bibnamefont {Zhu}}, \bibinfo {author} {\bibfnamefont {Marcel}\ \bibnamefont {Franz}}, \ and\ \bibinfo {author} {\bibfnamefont {Ian}\ \bibnamefont {Affleck}},\ }\bibfield  {title} {{\selectlanguage {en}\enquote {\bibinfo {title} {Emergent {Supersymmetry} from {Strongly} {Interacting} {Majorana} {Zero} {Modes}},}\ }}\href {\doibase 10.1103/PhysRevLett.115.166401} {\bibfield  {journal} {\bibinfo  {journal} {Physical Review Letters}\ }\textbf {\bibinfo {volume} {115}},\ \bibinfo {pages} {166401} (\bibinfo {year} {2015})}\BibitemShut {NoStop}%
\bibitem [{\citenamefont {Schollwöck}\ \emph {et~al.}(2004)\citenamefont {Schollwöck}, \citenamefont {Richter}, \citenamefont {Farnell},\ and\ \citenamefont {Bishop}}]{schollwock_quantum_2004}%
  \BibitemOpen
  \bibinfo {editor} {\bibfnamefont {Ulrich}\ \bibnamefont {Schollwöck}}, \bibinfo {editor} {\bibfnamefont {Johannes}\ \bibnamefont {Richter}}, \bibinfo {editor} {\bibfnamefont {Damian J.~J.}\ \bibnamefont {Farnell}}, \ and\ \bibinfo {editor} {\bibfnamefont {Raymod~F.}\ \bibnamefont {Bishop}},\ eds.,\ \href {\doibase 10.1007/b96825} {{\selectlanguage {en}\emph {\bibinfo {title} {Quantum {Magnetism}}}}},\ \bibinfo {series} {Lecture {Notes} in {Physics}}, Vol.\ \bibinfo {volume} {645}\ (\bibinfo  {publisher} {Springer Berlin Heidelberg},\ \bibinfo {address} {Berlin, Heidelberg},\ \bibinfo {year} {2004})\BibitemShut {NoStop}%
\bibitem [{\citenamefont {Mikeska}(1977)}]{mikeska_solitons_1977}%
  \BibitemOpen
  \bibfield  {author} {\bibinfo {author} {\bibfnamefont {H~J}\ \bibnamefont {Mikeska}},\ }\bibfield  {title} {{\selectlanguage {en}\enquote {\bibinfo {title} {Solitons in a one-dimensional magnet with an easy plane},}\ }}\href {\doibase 10.1088/0022-3719/11/1/007} {\bibfield  {journal} {\bibinfo  {journal} {Journal of Physics C: Solid State Physics}\ }\textbf {\bibinfo {volume} {11}},\ \bibinfo {pages} {L29--L32} (\bibinfo {year} {1977})}\BibitemShut {NoStop}%
\bibitem [{\citenamefont {Witten}(1978)}]{witten_properties_1978}%
  \BibitemOpen
  \bibfield  {author} {\bibinfo {author} {\bibfnamefont {Edward}\ \bibnamefont {Witten}},\ }\bibfield  {title} {{\selectlanguage {en}\enquote {\bibinfo {title} {Some properties of the {Psi}{\textasciicircum}2 model in two dimensions},}\ }}\href {https://doi.org/10.1016/0550-3213(78)90204-3} {\bibfield  {journal} {\bibinfo  {journal} {Nuclear Physics B}\ }\textbf {\bibinfo {volume} {142}},\ \bibinfo {pages} {285--300} (\bibinfo {year} {1978})}\BibitemShut {NoStop}%
\bibitem [{\citenamefont {Chen}(2019)}]{chen_notitle_2019}%
  \BibitemOpen
  \bibfield  {author} {\bibinfo {author} {\bibfnamefont {Chun}\ \bibnamefont {Chen}},\ }\href {http://arxiv.org/abs/1908.10777} {} (\bibinfo {year} {2019}),\ \bibinfo {note} {arXiv:1908.10777 [cond-mat]}\BibitemShut {NoStop}%
\bibitem [{\citenamefont {Mogi}\ \emph {et~al.}(2019)\citenamefont {Mogi}, \citenamefont {Nakajima}, \citenamefont {Ukleev}, \citenamefont {Tsukazaki}, \citenamefont {Yoshimi} \emph {et~al.}}]{mogi2019large}%
  \BibitemOpen
  \bibfield  {author} {\bibinfo {author} {\bibfnamefont {Masataka}\ \bibnamefont {Mogi}}, \bibinfo {author} {\bibfnamefont {Taro}\ \bibnamefont {Nakajima}}, \bibinfo {author} {\bibfnamefont {Victor}\ \bibnamefont {Ukleev}}, \bibinfo {author} {\bibfnamefont {Atsushi}\ \bibnamefont {Tsukazaki}}, \bibinfo {author} {\bibfnamefont {Ryutaro}\ \bibnamefont {Yoshimi}},  \emph {et~al.},\ }\bibfield  {title} {\enquote {\bibinfo {title} {Large anomalous hall effect in topological insulators with proximitized ferromagnetic insulators},}\ }\href@noop {} {\bibfield  {journal} {\bibinfo  {journal} {Physical review letters}\ }\textbf {\bibinfo {volume} {123}},\ \bibinfo {pages} {016804} (\bibinfo {year} {2019})}\BibitemShut {NoStop}%
\bibitem [{\citenamefont {Watanabe}\ \emph {et~al.}(2019)\citenamefont {Watanabe}, \citenamefont {Yoshimi}, \citenamefont {Kawamura}, \citenamefont {Mogi}, \citenamefont {Tsukazaki}, \citenamefont {Yu}, \citenamefont {Nakajima}, \citenamefont {Takahashi}, \citenamefont {Kawasaki},\ and\ \citenamefont {Tokura}}]{watanabe2019}%
  \BibitemOpen
  \bibfield  {author} {\bibinfo {author} {\bibfnamefont {R.}~\bibnamefont {Watanabe}}, \bibinfo {author} {\bibfnamefont {R.}~\bibnamefont {Yoshimi}}, \bibinfo {author} {\bibfnamefont {M.}~\bibnamefont {Kawamura}}, \bibinfo {author} {\bibfnamefont {M.}~\bibnamefont {Mogi}}, \bibinfo {author} {\bibfnamefont {A.}~\bibnamefont {Tsukazaki}}, \bibinfo {author} {\bibfnamefont {X.~Z.}\ \bibnamefont {Yu}}, \bibinfo {author} {\bibfnamefont {K.}~\bibnamefont {Nakajima}}, \bibinfo {author} {\bibfnamefont {K.~S.}\ \bibnamefont {Takahashi}}, \bibinfo {author} {\bibfnamefont {M.}~\bibnamefont {Kawasaki}}, \ and\ \bibinfo {author} {\bibfnamefont {Y.}~\bibnamefont {Tokura}},\ }\bibfield  {title} {\enquote {\bibinfo {title} {Quantum anomalous hall effect driven by magnetic proximity coupling in all-telluride based heterostructure},}\ }\href {\doibase 10.1063/1.5111891} {\bibfield  {journal} {\bibinfo  {journal} {Applied Physics Letters}\ }\textbf {\bibinfo {volume} {115}},\ \bibinfo {pages} {102403} (\bibinfo {year} {2019})},\
  \Eprint {http://arxiv.org/abs/https://pubs.aip.org/aip/apl/article-pdf/doi/10.1063/1.5111891/13270032/102403\_1\_online.pdf} {https://pubs.aip.org/aip/apl/article-pdf/doi/10.1063/1.5111891/13270032/102403\_1\_online.pdf} \BibitemShut {NoStop}%
\bibitem [{\citenamefont {Xia}\ \emph {et~al.}(2009)\citenamefont {Xia}, \citenamefont {Qian}, \citenamefont {Hsieh}, \citenamefont {Wray}, \citenamefont {Pal}, \citenamefont {Lin}, \citenamefont {Bansil}, \citenamefont {Grauer}, \citenamefont {Hor}, \citenamefont {Cava},\ and\ \citenamefont {Hasan}}]{xia_observation_2009}%
  \BibitemOpen
  \bibfield  {author} {\bibinfo {author} {\bibfnamefont {Y.}~\bibnamefont {Xia}}, \bibinfo {author} {\bibfnamefont {D.}~\bibnamefont {Qian}}, \bibinfo {author} {\bibfnamefont {D.}~\bibnamefont {Hsieh}}, \bibinfo {author} {\bibfnamefont {L.}~\bibnamefont {Wray}}, \bibinfo {author} {\bibfnamefont {A.}~\bibnamefont {Pal}}, \bibinfo {author} {\bibfnamefont {H.}~\bibnamefont {Lin}}, \bibinfo {author} {\bibfnamefont {A.}~\bibnamefont {Bansil}}, \bibinfo {author} {\bibfnamefont {D.}~\bibnamefont {Grauer}}, \bibinfo {author} {\bibfnamefont {Y.~S.}\ \bibnamefont {Hor}}, \bibinfo {author} {\bibfnamefont {R.~J.}\ \bibnamefont {Cava}}, \ and\ \bibinfo {author} {\bibfnamefont {M.~Z.}\ \bibnamefont {Hasan}},\ }\bibfield  {title} {\enquote {\bibinfo {title} {Observation of a large-gap topological-insulator class with a single {Dirac} cone on the surface},}\ }\href {\doibase 10.1038/nphys1274} {\bibfield  {journal} {\bibinfo  {journal} {Nature Physics}\ }\textbf {\bibinfo {volume} {5}},\ \bibinfo {pages} {398--402} (\bibinfo
  {year} {2009})}\BibitemShut {NoStop}%
\bibitem [{\citenamefont {Zhang}\ \emph {et~al.}(2009)\citenamefont {Zhang}, \citenamefont {Liu}, \citenamefont {Qi}, \citenamefont {Dai}, \citenamefont {Fang},\ and\ \citenamefont {Zhang}}]{zhang_topological_2009}%
  \BibitemOpen
  \bibfield  {author} {\bibinfo {author} {\bibfnamefont {Haijun}\ \bibnamefont {Zhang}}, \bibinfo {author} {\bibfnamefont {Chao-Xing}\ \bibnamefont {Liu}}, \bibinfo {author} {\bibfnamefont {Xiao-Liang}\ \bibnamefont {Qi}}, \bibinfo {author} {\bibfnamefont {Xi}~\bibnamefont {Dai}}, \bibinfo {author} {\bibfnamefont {Zhong}\ \bibnamefont {Fang}}, \ and\ \bibinfo {author} {\bibfnamefont {Shou-Cheng}\ \bibnamefont {Zhang}},\ }\bibfield  {title} {\enquote {\bibinfo {title} {Topological insulators in {Bi2Se3}, {Bi2Te3} and {Sb2Te3} with a single {Dirac} cone on the surface},}\ }\href {\doibase 10.1038/nphys1270} {\bibfield  {journal} {\bibinfo  {journal} {Nature Physics}\ }\textbf {\bibinfo {volume} {5}},\ \bibinfo {pages} {438--442} (\bibinfo {year} {2009})}\BibitemShut {NoStop}%
\bibitem [{\citenamefont {Popova}\ \emph {et~al.}(2013)\citenamefont {Popova}, \citenamefont {{Franco Galeano}}, \citenamefont {Deb}, \citenamefont {Warot-Fonrose}, \citenamefont {Kachkachi}, \citenamefont {Gendron}, \citenamefont {Ott}, \citenamefont {Berini},\ and\ \citenamefont {Keller}}]{POPOVA2013139}%
  \BibitemOpen
  \bibfield  {author} {\bibinfo {author} {\bibfnamefont {Elena}\ \bibnamefont {Popova}}, \bibinfo {author} {\bibfnamefont {Andres~Felipe}\ \bibnamefont {{Franco Galeano}}}, \bibinfo {author} {\bibfnamefont {Marwan}\ \bibnamefont {Deb}}, \bibinfo {author} {\bibfnamefont {Bénédicte}\ \bibnamefont {Warot-Fonrose}}, \bibinfo {author} {\bibfnamefont {Hamid}\ \bibnamefont {Kachkachi}}, \bibinfo {author} {\bibfnamefont {François}\ \bibnamefont {Gendron}}, \bibinfo {author} {\bibfnamefont {Frédéric}\ \bibnamefont {Ott}}, \bibinfo {author} {\bibfnamefont {Bruno}\ \bibnamefont {Berini}}, \ and\ \bibinfo {author} {\bibfnamefont {Niels}\ \bibnamefont {Keller}},\ }\bibfield  {title} {\enquote {\bibinfo {title} {Magnetic anisotropies in ultrathin bismuth iron garnet films},}\ }\href {\doibase https://doi.org/10.1016/j.jmmm.2013.02.003} {\bibfield  {journal} {\bibinfo  {journal} {Journal of Magnetism and Magnetic Materials}\ }\textbf {\bibinfo {volume} {335}},\ \bibinfo {pages} {139--143} (\bibinfo {year}
  {2013})}\BibitemShut {NoStop}%
\bibitem [{\citenamefont {Zhang}\ \emph {et~al.}(2016)\citenamefont {Zhang}, \citenamefont {Zhao}, \citenamefont {Song}, \citenamefont {Jia}, \citenamefont {Shi},\ and\ \citenamefont {Han}}]{Zhang_2016}%
  \BibitemOpen
  \bibfield  {author} {\bibinfo {author} {\bibfnamefont {Xiao}\ \bibnamefont {Zhang}}, \bibinfo {author} {\bibfnamefont {Yuelei}\ \bibnamefont {Zhao}}, \bibinfo {author} {\bibfnamefont {Qi}~\bibnamefont {Song}}, \bibinfo {author} {\bibfnamefont {Shuang}\ \bibnamefont {Jia}}, \bibinfo {author} {\bibfnamefont {Jing}\ \bibnamefont {Shi}}, \ and\ \bibinfo {author} {\bibfnamefont {Wei}\ \bibnamefont {Han}},\ }\bibfield  {title} {\enquote {\bibinfo {title} {Magnetic anisotropy of the single-crystalline ferromagnetic insulator {C}r2{G}e2{T}e6},}\ }\href {\doibase 10.7567/JJAP.55.033001} {\bibfield  {journal} {\bibinfo  {journal} {Japanese Journal of Applied Physics}\ }\textbf {\bibinfo {volume} {55}},\ \bibinfo {pages} {033001} (\bibinfo {year} {2016})}\BibitemShut {NoStop}%
\bibitem [{\citenamefont {Talalaevskij}\ \emph {et~al.}(2017)\citenamefont {Talalaevskij}, \citenamefont {Decker}, \citenamefont {Stigloher}, \citenamefont {Mitra}, \citenamefont {K{\"o}rner}, \citenamefont {Cespedes}, \citenamefont {Back},\ and\ \citenamefont {Hickey}}]{talalaevskij2017magnetic}%
  \BibitemOpen
  \bibfield  {author} {\bibinfo {author} {\bibfnamefont {A}~\bibnamefont {Talalaevskij}}, \bibinfo {author} {\bibfnamefont {M}~\bibnamefont {Decker}}, \bibinfo {author} {\bibfnamefont {J}~\bibnamefont {Stigloher}}, \bibinfo {author} {\bibfnamefont {A}~\bibnamefont {Mitra}}, \bibinfo {author} {\bibfnamefont {HS}~\bibnamefont {K{\"o}rner}}, \bibinfo {author} {\bibfnamefont {O}~\bibnamefont {Cespedes}}, \bibinfo {author} {\bibfnamefont {CH}~\bibnamefont {Back}}, \ and\ \bibinfo {author} {\bibfnamefont {BJ}~\bibnamefont {Hickey}},\ }\bibfield  {title} {\enquote {\bibinfo {title} {Magnetic properties of spin waves in thin yttrium iron garnet films},}\ }\href@noop {} {\bibfield  {journal} {\bibinfo  {journal} {Physical Review B}\ }\textbf {\bibinfo {volume} {95}},\ \bibinfo {pages} {064409} (\bibinfo {year} {2017})}\BibitemShut {NoStop}%
\bibitem [{\citenamefont {Mogi}\ \emph {et~al.}(2018)\citenamefont {Mogi}, \citenamefont {Tsukazaki}, \citenamefont {Kaneko}, \citenamefont {Yoshimi}, \citenamefont {Takahashi}, \citenamefont {Kawasaki},\ and\ \citenamefont {Tokura}}]{mogi2018}%
  \BibitemOpen
  \bibfield  {author} {\bibinfo {author} {\bibfnamefont {M.}~\bibnamefont {Mogi}}, \bibinfo {author} {\bibfnamefont {A.}~\bibnamefont {Tsukazaki}}, \bibinfo {author} {\bibfnamefont {Y.}~\bibnamefont {Kaneko}}, \bibinfo {author} {\bibfnamefont {R.}~\bibnamefont {Yoshimi}}, \bibinfo {author} {\bibfnamefont {K.~S.}\ \bibnamefont {Takahashi}}, \bibinfo {author} {\bibfnamefont {M.}~\bibnamefont {Kawasaki}}, \ and\ \bibinfo {author} {\bibfnamefont {Y.}~\bibnamefont {Tokura}},\ }\bibfield  {title} {\enquote {\bibinfo {title} {Ferromagnetic insulator {C}r2{G}e2{T}e6 thin films with perpendicular remanence},}\ }\href@noop {} {\bibfield  {journal} {\bibinfo  {journal} {APL Materials}\ }\textbf {\bibinfo {volume} {6}},\ \bibinfo {pages} {091104} (\bibinfo {year} {2018})}\BibitemShut {NoStop}%
\bibitem [{\citenamefont {Tian}\ \emph {et~al.}(2021)\citenamefont {Tian}, \citenamefont {Pan}, \citenamefont {Wang}, \citenamefont {Ye}, \citenamefont {Sheng}, \citenamefont {Wang}, \citenamefont {Liu}, \citenamefont {Huang}, \citenamefont {Zhang}, \citenamefont {Xu} \emph {et~al.}}]{tian2021dyocl}%
  \BibitemOpen
  \bibfield  {author} {\bibinfo {author} {\bibfnamefont {Congkuan}\ \bibnamefont {Tian}}, \bibinfo {author} {\bibfnamefont {Feihao}\ \bibnamefont {Pan}}, \bibinfo {author} {\bibfnamefont {Le}~\bibnamefont {Wang}}, \bibinfo {author} {\bibfnamefont {Dehua}\ \bibnamefont {Ye}}, \bibinfo {author} {\bibfnamefont {Jieming}\ \bibnamefont {Sheng}}, \bibinfo {author} {\bibfnamefont {Jinchen}\ \bibnamefont {Wang}}, \bibinfo {author} {\bibfnamefont {Juanjuan}\ \bibnamefont {Liu}}, \bibinfo {author} {\bibfnamefont {Jiale}\ \bibnamefont {Huang}}, \bibinfo {author} {\bibfnamefont {Hongxia}\ \bibnamefont {Zhang}}, \bibinfo {author} {\bibfnamefont {Daye}\ \bibnamefont {Xu}},  \emph {et~al.},\ }\bibfield  {title} {\enquote {\bibinfo {title} {Dy{O}{C}l: A rare-earth based two-dimensional van der {W}aals material with strong magnetic anisotropy},}\ }\href@noop {} {\bibfield  {journal} {\bibinfo  {journal} {Physical Review B}\ }\textbf {\bibinfo {volume} {104}},\ \bibinfo {pages} {214410} (\bibinfo {year} {2021})}\BibitemShut
  {NoStop}%
\bibitem [{\citenamefont {Tian}\ \emph {et~al.}(2016)\citenamefont {Tian}, \citenamefont {Gray}, \citenamefont {Ji}, \citenamefont {Cava},\ and\ \citenamefont {Burch}}]{Tian_2016}%
  \BibitemOpen
  \bibfield  {author} {\bibinfo {author} {\bibfnamefont {Yao}\ \bibnamefont {Tian}}, \bibinfo {author} {\bibfnamefont {Mason~J}\ \bibnamefont {Gray}}, \bibinfo {author} {\bibfnamefont {Huiwen}\ \bibnamefont {Ji}}, \bibinfo {author} {\bibfnamefont {R~J}\ \bibnamefont {Cava}}, \ and\ \bibinfo {author} {\bibfnamefont {Kenneth~S}\ \bibnamefont {Burch}},\ }\bibfield  {title} {\enquote {\bibinfo {title} {Magneto-elastic coupling in a potential ferromagnetic 2d atomic crystal},}\ }\href {\doibase 10.1088/2053-1583/3/2/025035} {\bibfield  {journal} {\bibinfo  {journal} {2D Materials}\ }\textbf {\bibinfo {volume} {3}},\ \bibinfo {pages} {025035} (\bibinfo {year} {2016})}\BibitemShut {NoStop}%
\bibitem [{\citenamefont {Wang}\ \emph {et~al.}(2016)\citenamefont {Wang}, \citenamefont {Du}, \citenamefont {Fredrik~Liu}, \citenamefont {Hu}, \citenamefont {Zhang}, \citenamefont {Zhang}, \citenamefont {Owen} \emph {et~al.}}]{Wang_2016}%
  \BibitemOpen
  \bibfield  {author} {\bibinfo {author} {\bibfnamefont {Xingzhi}\ \bibnamefont {Wang}}, \bibinfo {author} {\bibfnamefont {Kezhao}\ \bibnamefont {Du}}, \bibinfo {author} {\bibfnamefont {Yu~Yang}\ \bibnamefont {Fredrik~Liu}}, \bibinfo {author} {\bibfnamefont {Peng}\ \bibnamefont {Hu}}, \bibinfo {author} {\bibfnamefont {Jun}\ \bibnamefont {Zhang}}, \bibinfo {author} {\bibfnamefont {Qing}\ \bibnamefont {Zhang}}, \bibinfo {author} {\bibfnamefont {Man Hon~Samuel}\ \bibnamefont {Owen}},  \emph {et~al.},\ }\bibfield  {title} {\enquote {\bibinfo {title} {Raman spectroscopy of atomically thin two-dimensional magnetic iron phosphorus trisulfide (feps3) crystals},}\ }\href {\doibase 10.1088/2053-1583/3/3/031009} {\bibfield  {journal} {\bibinfo  {journal} {2D Materials}\ }\textbf {\bibinfo {volume} {3}},\ \bibinfo {pages} {031009} (\bibinfo {year} {2016})}\BibitemShut {NoStop}%
\bibitem [{\citenamefont {Huang}\ \emph {et~al.}(2017)\citenamefont {Huang}, \citenamefont {Clark}, \citenamefont {Navarro-Moratalla}, \citenamefont {Klein}, \citenamefont {Cheng}, \citenamefont {Seyler}, \citenamefont {Zhong}, \citenamefont {Schmidgall}, \citenamefont {McGuire}, \citenamefont {Cobden} \emph {et~al.}}]{huang2017layer}%
  \BibitemOpen
  \bibfield  {author} {\bibinfo {author} {\bibfnamefont {Bevin}\ \bibnamefont {Huang}}, \bibinfo {author} {\bibfnamefont {Genevieve}\ \bibnamefont {Clark}}, \bibinfo {author} {\bibfnamefont {Efr{\'e}n}\ \bibnamefont {Navarro-Moratalla}}, \bibinfo {author} {\bibfnamefont {Dahlia~R}\ \bibnamefont {Klein}}, \bibinfo {author} {\bibfnamefont {Ran}\ \bibnamefont {Cheng}}, \bibinfo {author} {\bibfnamefont {Kyle~L}\ \bibnamefont {Seyler}}, \bibinfo {author} {\bibfnamefont {Ding}\ \bibnamefont {Zhong}}, \bibinfo {author} {\bibfnamefont {Emma}\ \bibnamefont {Schmidgall}}, \bibinfo {author} {\bibfnamefont {Michael~A}\ \bibnamefont {McGuire}}, \bibinfo {author} {\bibfnamefont {David~H}\ \bibnamefont {Cobden}},  \emph {et~al.},\ }\bibfield  {title} {\enquote {\bibinfo {title} {Layer-dependent ferromagnetism in a van der waals crystal down to the monolayer limit},}\ }\href@noop {} {\bibfield  {journal} {\bibinfo  {journal} {Nature}\ }\textbf {\bibinfo {volume} {546}},\ \bibinfo {pages} {270--273} (\bibinfo {year}
  {2017})}\BibitemShut {NoStop}%
\bibitem [{\citenamefont {Huijse}\ \emph {et~al.}(2015)\citenamefont {Huijse}, \citenamefont {Bauer},\ and\ \citenamefont {Berg}}]{huijse_emergent_2015}%
  \BibitemOpen
  \bibfield  {author} {\bibinfo {author} {\bibfnamefont {Liza}\ \bibnamefont {Huijse}}, \bibinfo {author} {\bibfnamefont {Bela}\ \bibnamefont {Bauer}}, \ and\ \bibinfo {author} {\bibfnamefont {Erez}\ \bibnamefont {Berg}},\ }\bibfield  {title} {{\selectlanguage {en}\enquote {\bibinfo {title} {Emergent {Supersymmetry} at the {Ising}–{Berezinskii}-{Kosterlitz}-{Thouless} {Multicritical} {Point}},}\ }}\href {\doibase 10.1103/PhysRevLett.114.090404} {\bibfield  {journal} {\bibinfo  {journal} {Physical Review Letters}\ }\textbf {\bibinfo {volume} {114}},\ \bibinfo {pages} {090404} (\bibinfo {year} {2015})}\BibitemShut {NoStop}%
\bibitem [{\citenamefont {Gogolin}\ \emph {et~al.}(2004)\citenamefont {Gogolin}, \citenamefont {Nersesyan},\ and\ \citenamefont {Tsvelik}}]{gogolin_bosonization_2004}%
  \BibitemOpen
  \bibfield  {author} {\bibinfo {author} {\bibfnamefont {Alexander~O.}\ \bibnamefont {Gogolin}}, \bibinfo {author} {\bibfnamefont {Alexander~A.}\ \bibnamefont {Nersesyan}}, \ and\ \bibinfo {author} {\bibfnamefont {Alexei~M.}\ \bibnamefont {Tsvelik}},\ }\href@noop {} {{\selectlanguage {eng}\emph {\bibinfo {title} {Bosonization and strongly correlated systems}}}}\ (\bibinfo  {publisher} {Cambridge University Press},\ \bibinfo {address} {Cambridge},\ \bibinfo {year} {2004})\BibitemShut {NoStop}%
\bibitem [{\citenamefont {Giamarchi}(2004)}]{giamarchi_quantum_2004}%
  \BibitemOpen
  \bibfield  {author} {\bibinfo {author} {\bibfnamefont {Thierry}\ \bibnamefont {Giamarchi}},\ }\href@noop {} {\emph {\bibinfo {title} {Quantum physics in one dimension}}}\ (\bibinfo  {publisher} {Clarendon Press},\ \bibinfo {address} {Oxford},\ \bibinfo {year} {2004})\BibitemShut {NoStop}%
\end{thebibliography}%

\end{document}